%% file: sample-manuscript.tex
\documentclass[acmsmall]{acmart}

\AtBeginDocument{%
  \providecommand\BibTeX{{%
    \normalfont B\kern-0.5em{\scshape i\kern-0.25em b}\kern-0.8em\TeX}}}

\setcopyright{acmlicensed}
\acmJournal{TOSEM}
\acmYear{2024}

\usepackage{multirow}
\usepackage{arydshln}
\usepackage{subcaption}
\usepackage{colortbl}
\usepackage{amsfonts}
\usepackage{tcolorbox}
\usepackage{algorithm}
\usepackage{algorithmicx}
\usepackage{algpseudocode}

\algnewcommand\algorithmicinitialize{\textbf{Initialize}}
\algnewcommand\Initialize{\item[\algorithmicinitialize]}
\algnewcommand\algorithmicFor{\textbf{For}}
\algnewcommand\for{\item[\algorithmicFor]}

\newcommand{\primarystudies}{57}

\begin{document}

\title{A Systematic Literature Review on Neural Code Translation}

\author{Xiang Chen}
\authornote{Xiang Chen and Jiacheng Xue contributed equally to this work and are recognized as co-first authors.}
\authornote{Xiang Chen is the corresponding author.}
\affiliation{%
    \institution{School of Artificial Intelligence and Computer Science, Nantong University}
    \city{Nantong}
  \country{China}
}
\affiliation{%
    \institution{State Key Lab. for Novel Software Technology, Nanjing University}
    \city{Nanjing}
  \country{China}
}
\email{xchencs@ntu.edu.cn}

\author{Jiacheng Xue}
\affiliation{%
    \institution{School of Artificial Intelligence and Computer Science, Nantong University}
    \city{Nantong}
  \country{China}
}
\email{xjc202603@gmail.com}

\author{Xiaofei Xie}
\affiliation{%
    \institution{ School of Computing and Information Systems, Singapore Management University}   
  \country{Singapore}
}
\email{fxie@smu.edu.sg}

\author{Caokai Liang}
\affiliation{%
    \institution{School of Artificial Intelligence and Computer Science, Nantong University}
    \city{Nantong}
  \country{China}
}
\email{Lcaokai@stmail.ntu.edu.cn}

\author{Xiaolin Ju}
\affiliation{%
    \institution{School of Artificial Intelligence and Computer Science, Nantong University}
    \city{Nantong}
  \country{China}
}
\email{ju.xl@ntu.edu.cn}

\renewcommand{\shortauthors}{Chen, et al.}

\begin{abstract}
Code translation aims to convert code from one programming language to another automatically. It is motivated by the need for multi-language software development and legacy system migration. 
In recent years, neural code translation has gained significant attention, driven by rapid advancements in deep learning and large language models. 
Researchers have proposed various techniques to improve neural code translation quality. However, to the best of our knowledge, no comprehensive systematic literature review has
been conducted to summarize the key techniques and challenges in this field.
To fill this research gap, we collected {\primarystudies} primary studies covering the period 2020$\sim$2025 on neural code translation. These studies are analyzed from seven key perspectives: task characteristics, data preprocessing, code modeling, model construction, post-processing, evaluation subjects, and evaluation metrics. Our analysis reveals current research trends, identifies unresolved challenges, and shows potential directions for future work. These findings can provide valuable insights for both researchers and practitioners in the field of neural code translation.
\end{abstract}



\maketitle

\input{sections/1.introduction}

\input{sections/2.background}

\input{sections/3.methodology}

\input{sections/4.resultanalysis}
\input{sections/5.open}

\input{sections/6.threat}
\input{sections/7.conclusion}

\begin{acks}
Jiacheng Xue and Xiang Chen have contributed equally to this work, and they are co-first authors. Xiang Chen is the corresponding author. 
This research was partially supported by the National Natural Science Foundation of China (Grant No. 61202006), the Open Project of State Key Laboratory for Novel Software Technology at Nanjing University (Grant No. KFKT2024B21), and the Postgraduate Research \& Practice Innovation Program of Jiangsu Province (Grant No. SJCX25\_2005).
\end{acks}

\bibliographystyle{ACM-Reference-Format}
\bibliography{sample-base}

\appendix

\end{document}

%% file: sections/1.introduction.tex
\section{Introduction}

Code translation aims to automatically convert code written in one programming language into another. This task is becoming increasingly important in the field of software engineering, driven primarily by the growing complexity of software systems and the rising demand for multi-language development. As modern applications are often deployed across multiple platforms, developers frequently need to implement or maintain the same functionality in different programming languages to ensure cross-platform compatibility and enhance software accessibility~\cite{tao2024unraveling}.

In large-scale software development, it is common practice for programmers to use multiple programming languages rather than relying exclusively on a single one~\cite{chisnall2013challenge}. This polyglot programming paradigm enables developers to leverage the strengths of different languages for specific tasks, incrementally migrate existing systems across languages, and reuse legacy codebases~\cite{grimmer2018cross}. Among these practices, code migration plays a crucial role, as it not only supports the adaptation of legacy systems to modern platforms but also significantly reduces long-term maintenance costs~\cite{zhu2024semi}.
As more companies are required to migrate their project code from legacy programming languages to newer ones that offer enhanced features and align with modern developer practices, the demand for high-quality program translation is steadily increasing. For example, the Commonwealth Bank of Australia migrated its platform from COBOL to Java~\cite{zhong2010mining}. Migrating these legacy systems provides greater flexibility, better system understanding, easier maintenance, and cost reduction~\cite{roziere2021leveraging}.
In addition to supporting cross-language development and legacy system migration, automated code translation can significantly reduce system maintenance costs. It also helps improve various non-functional properties of software systems, including security, scalability, and performance. As a result, automated code translation has emerged as a vital research focus within the field of software engineering.

In the early stages of code translation research, approaches predominantly relied on manually crafted rules~\cite{khalafinejad2010rule,emani2017dbridge,noor2020translation}. These rule-based approaches typically utilized techniques (such as pattern matching and syntax tree transformations) to map the syntactic structures of source code to those of the target language. Although they could achieve high accuracy within specific domains or language pairs, they heavily depended on expert knowledge for rule design and ongoing maintenance. Moreover, these approaches often faced significant challenges in terms of scalability, incurred high maintenance costs, and were prone to errors when applied to new programming languages or source code with complex logic.

With the advancement of machine learning, a growing number of studies have adopted supervised and unsupervised learning approaches for code translation. In particular, supervised learning utilizes existing parallel corpora to generate code that better aligns with real-world programming styles. This paradigm is supported by techniques such as statistical machine translation~\cite{nguyen2015divide} and tree-based neural networks~\cite{chen2018tree}. However, its performance and generalizability are often limited by the availability and quality of parallel corpora.
To mitigate this limitation, unsupervised learning has emerged as a promising alternative. Roziere et al.~\cite{roziere2020unsupervised} introduced TransCoder, which leverages back-translation to train models without relying on parallel corpora. The following studies~\cite{roziere2021leveraging,szafraniec2022code} further improved translation quality by incorporating unit tests and compiler representations into the training process. However, these approaches still require language-specific training for different programming language pairs, which results in high computational costs and limited scalability.

The emergence of the deep learning era marked a significant shift in the landscape of code translation. In particular, the introduction of the Transformer model~\cite{vaswani2017attention} greatly accelerated progress in this field. Its self-attention mechanism not only enabled better modeling of long-range dependencies but also significantly improved training and inference efficiency, thereby enhancing the model’s ability to understand code structure and semantics.
As a result, researchers began modeling the code translation task as a Neural Machine Translation (NMT) problem and proposed targeted approaches~\cite{roziere2020unsupervised,chen2018tree}. With the rapid growth of data availability and computational resources, pre-trained models (such as CodeT5~\cite{wang2021codet5}) were adopted to automatically learn code translation rules from large-scale parallel corpora in an end-to-end way.
In this survey, we refer to research problems based on such approaches as neural code translation.

However, research in neural code translation faces several key challenges.
First, the scarcity of parallel code corpora limits the ability of translation models to learn precise cross-language alignments~\cite{zhu2024semi}. Without sufficient training data, models may struggle to establish accurate mappings between source and target programming languages, resulting in suboptimal translation quality.
Second, substantial differences in syntax and semantic structures across programming languages further complicate the translation process. For instance, translating between Python's dynamic type system and C++’s static type system requires careful handling of complex aspects such as type inference and memory management. Failure to account for these differences can lead to redundant or inefficient translated code, and in some cases, may even introduce runtime errors.
Third, the lack of transparency in model-generated outputs makes it difficult for developers to understand the reasoning behind translation decisions. In many cases, models operate primarily as token predictors, rather than truly understanding the semantics of the code translation task~\cite{abukhalaf2023codex,jiang2024self,white2023prompt,zhuo2023robustness}.
Finally, the absence of project-level information often results in translated code snippets that are incompatible with the broader context. Tasks requiring global program understanding (such as managing variable scopes, maintaining consistent naming conventions, or preserving the structure of nested loops) can be mishandled when models rely solely on local context.

Although many previous studies have explored neural code translation, to the best of our knowledge, no comprehensive Systematic Literature Review (SLR) has been conducted in this research topic.
To fill this gap, we selected {\primarystudies} primary studies between 2020 and 2025. 
Then we systematically analyzed previous studies from multiple perspectives, including task characteristics, data preprocessing, code modeling, model construction, post-processing, evaluation subjects, and evaluation metrics.
Finally, we identified the limitations of current research and proposed several practical future directions. 
By providing a comprehensive understanding of the state-of-the-art in neural code translation, our survey serves as a valuable resource for both practitioners and researchers to gain a comprehensive understanding of neural code translation and to advance future developments in the field.

The main contributions of our SLR can be summarized as follows:

\begin{itemize}
    \item We systematically collect and review {\primarystudies} primary studies published between 2020 and 2025 on neural code translation.
    \item Based on the collected studies, we present a comprehensive qualitative and quantitative synthesis from the following perspectives: task characteristics, data preprocessing, code modeling, model construction, post-processing, evaluation subjects, and evaluation metrics.
    \item We identify current limitations of neural code translation and provide practical implications for future studies.
\end{itemize}

The remainder of this SLR is organized as follows:
Section~\ref{sec:background} introduces the background and establishes the foundation for our SLR.
Section~\ref{sec:methodology} describes the review methodology adopted in our SLR.
Section~\ref{sec:result} answers the research questions and summarizes the main findings.
Section~\ref{sec:open} identifies open challenges and discusses potential directions for future research.
Section~\ref{sec:threat} examines the potential threats to the validity of our SLR.
Finally, Section~\ref{sec:conclusion} concludes the SLR by summarizing the key insights.

%% file: sections/2.background.tex
\section{Background}
\label{sec:background}

In this section, we first introduce the background of code translation. Given that most neural code translation studies model this task as a form of neural machine translation, we subsequently provide an overview of NMT to establish the necessary context.

\subsection{Code Translation}

In this SLR, we focus on code translation, which refers to the process of converting source code from one programming language to another while ensuring semantic equivalence. Specifically, given a source code snippet $X_i$ written in one programming language, the goal is to generate a corresponding translation $Y_i$ in a different language, ensuring that the original logic and behavior remain unchanged.

\begin{figure}[htbp] 
    \centering 
    \includegraphics[width=1.0\textwidth]{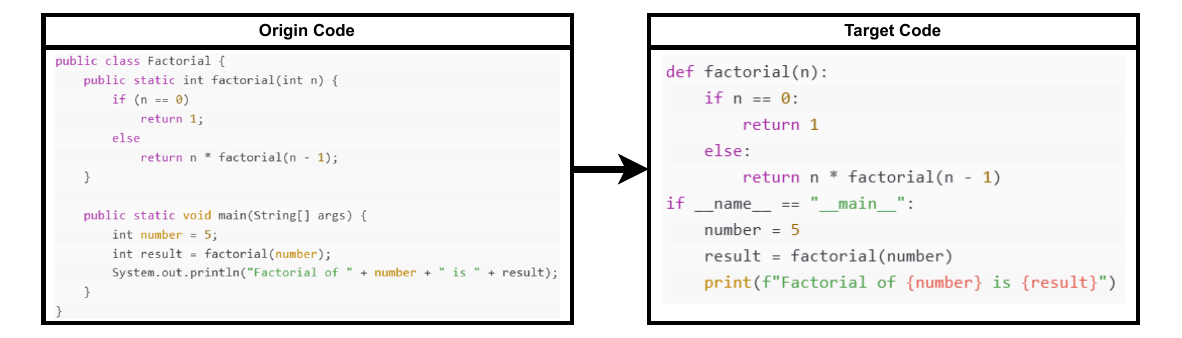} 
    \caption{Illustrative example of code translation from Java to Python} 
    \label{Fig:example} 
\end{figure}

We use the example shown in Fig.~\ref{Fig:example} to illustrate the translation of a Java program that computes the factorial into its Python equivalent. Although the syntax and structure of the two languages differ significantly, the core recursive logic remains unchanged. By understanding the fundamental constructs of both languages (such as classes and methods, function definitions, entry points, and output statements), the model should be able to perform effective code translation while preserving the original functionality.

Code generation and code refactoring are two tasks most closely related to code translation. Specifically, code generation involves producing code from natural language descriptions or high-level design specifications, whereas code refactoring focuses on improving the structure, readability, and maintainability of code without altering its external behavior. In contrast, code translation is specifically concerned with converting code from one programming language to another while ensuring functional equivalence.

\subsection{Neural Machine Translation}

With the rapid advancement of deep learning technologies, the field of Natural Language Processing (NLP) has made groundbreaking progress, with Neural Machine Translation (NMT) emerging as the dominant approach for translation tasks~\cite{jiang2020natural,ahammad2024improved}.
NMT models are primarily based on an encoder-decoder architecture~\cite{cho2014properties}, where the encoder transforms input sequences into semantic representations, which the decoder then uses to generate the target output sequentially. 
Compared to traditional statistical machine translation methods~\cite{lopez2008statistical}, NMT provides superior contextual understanding and greater flexibility.
Specifically, introducing the attention mechanism in NMT models can significantly enhance translation quality. 
By dynamically focusing on relevant parts of the source sentence during translation, the attention mechanism effectively reduces information loss in longer sentences. 
This breakthrough has improved translation performance and provided valuable insights for a wide range of other sequence-to-sequence tasks.

In recent years, neural code translation has emerged as an exciting yet challenging research area, increasingly integrating core concepts from NMT~\cite{chen2018tree,roziere2020unsupervised,roziere2021leveraging,szafraniec2022code}. This task involves converting source code from one programming language to another, requiring both strict adherence to language syntax and precise preservation of program logic.
By leveraging the encoder-decoder architecture and attention mechanism inherent in NMT, researchers can better learn the mapping between source language and target language, thereby improving the quality and robustness of code translation.

%% file: sections/3.methodology.tex
\section{Review Methodology}
\label{sec:methodology}

\begin{figure}[htbp] 
    \centering 
    \includegraphics[width=0.9\textwidth]{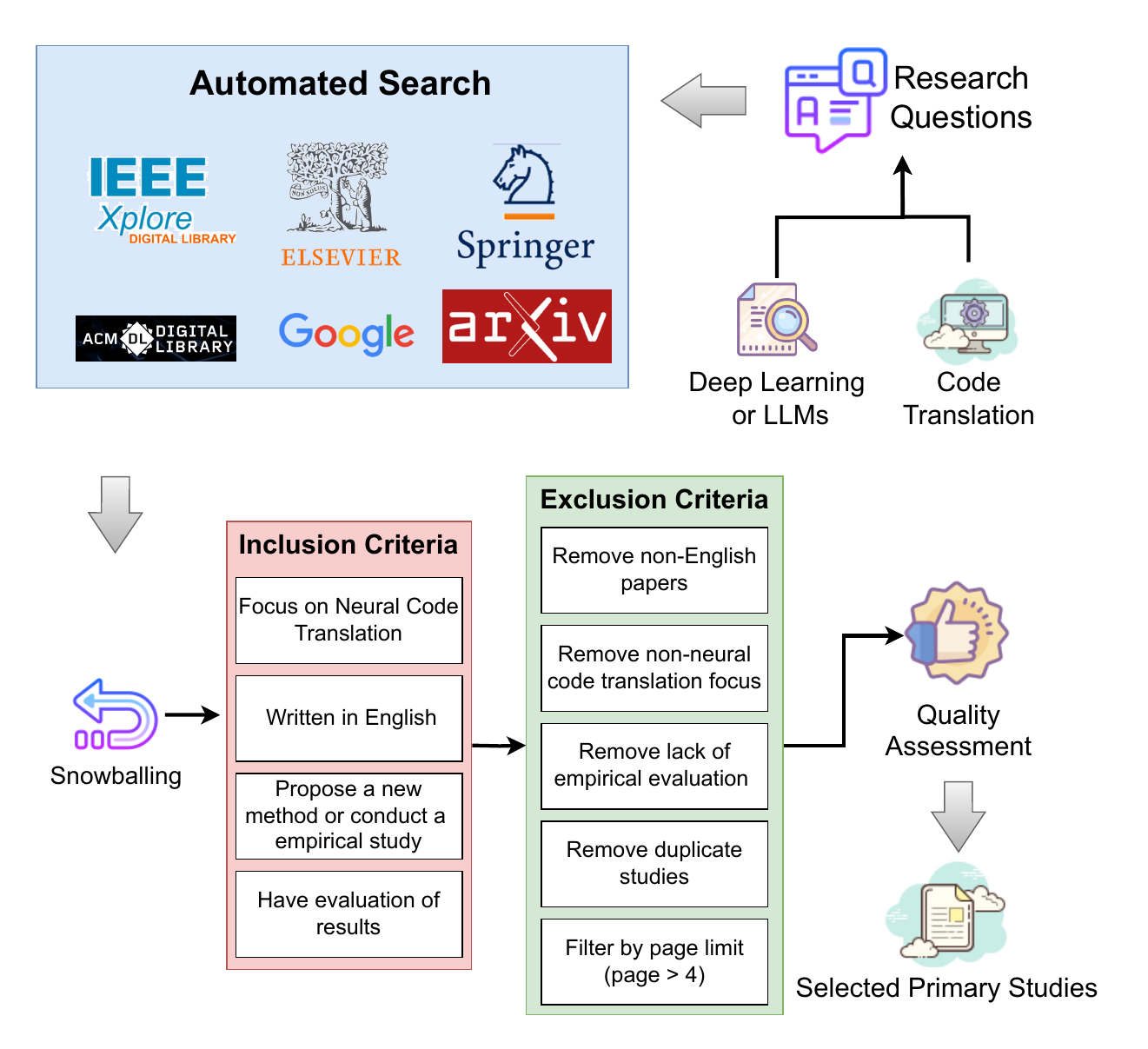} 
    \caption{Overview of the primary study search and collection process} 
    \label{Fig:collection} 
\end{figure}

Following the guidelines suggested by Kitchenham~\cite{kitchenham2009systematic}, we adopt the methodology shown in Fig.~\ref{Fig:collection} to search and select primary studies of neural code translation.

\subsection{Research Questions}

In our SLR, we aim to answer the following seven research questions (RQs), each designed to examine the progress of neural code translation research from different perspectives.

\textbf{RQ1: What are the task characteristics of neural code translation?}

\textbf{Motivation}: 
In RQ1, we aim to identify the key characteristics of the neural code translation task. Specifically, we focus on two main aspects: programming language pairs (such as C++ → Rust, and Java → Python) and code granularity (such as function-level and class-level).
By analyzing these aspects, we can identify research trends in language pair selection and code granularity settings.

\textbf{RQ2: What data preprocessing methods are used for neural code translation?}

\textbf{Motivation}: 
In RQ2, we aim to investigate the commonly used pre-processing methods for neural code translation. Pre-processing techniques (such as data cleaning, data deduplication, and data augmentation) play a crucial role in improving the quality of training data, which can significantly enhance the performance of neural code translation models.

\textbf{RQ3: What code modeling methods are used for neural code translation?}

\textbf{Motivation}: 
In RQ3, we aim to explore the common code modeling methods used in neural code translation. Based on whether structural information is considered, existing code modeling methods can be simply classified into three categories: text-based, graph-based, and IR-based methods. Among these, graph-based methods can be further divided into Abstract Syntax Trees (ASTs), Control Flow Graphs (CFGs), Data Flow Graphs (DFGs), and other representations. This RQ focuses on how different code modeling methods capture the syntactic and semantic information of code and examines the impact of structural representations on code translation quality.

\textbf{RQ4: What model construction methods are used for neural code translation?}

\textbf{Motivation}: 
In RQ4, we aim to explore the different model construction methods used for neural code translation, 
including training from scratch, fine-tuning, prompt engineering, multi-agent collaboration, and retrieval-augmented
generation.
By analyzing these methods, we aim to evaluate their respective strengths, limitations, and suitable application scenarios, thereby providing insights into their effectiveness across diverse contexts.

\textbf{RQ5: What post-processing methods are used for neural code translation?}

\textbf{Motivation}: 
In RQ5, we aim to investigate the post-processing methods used in neural code translation, which play a key role in improving the executability and readability of the translated code. Specifically, we analyze the proportion of studies that incorporate post-processing methods and categorize these methods into various types, 
including dynamic-based bug detection, static-based bug detection, bug repair, and code optimization. 

\textbf{RQ6: What are the evaluation subjects for neural code translation tasks?}

\textbf{Motivation}: 
In RQ6, we aim to provide a comprehensive summary of the evaluation subjects used in previous neural code translation studies. Specifically, we mainly focus on the sources of the experimental subjects, the covered programming language pairs, the code granularity, and the main measures taken to ensure the quality of the experimental subjects.

\textbf{RQ7: What evaluation metrics are used for neural code translation tasks?}

\textbf{Motivation}: 
In RQ7, we aim to examine the evaluation metrics used in previous neural code translation studies. 
We categorize these metrics into two main types: static-based and dynamic-based. 
Specifically, static-based metrics evaluate translated code without execution, measuring the similarity between the translated code and the ground truth. Common metrics include BLEU, CodeBLEU, and Exact Match (EM).
Dynamic-based metrics assess translation quality by executing the code and validating correctness through test cases. A typical metric is pass@k, which measures the test case pass rate.

\subsection{Search Strategy}

To conduct a comprehensive SLR on neural code translation, it is crucial to identify high-quality primary studies from academic databases. The chosen databases should cover a broad spectrum of research fields, including software engineering, machine learning, and natural language processing.
In this subsection, we present the list of academic databases selected for this SLR.

\begin{itemize}
    \item \textbf{IEEE Xplore.} This dataset provides access to top conferences and journals, such as \textit{IEEE Transactions on Software Engineering}.
    \item \textbf{ACM Digital Library.} This dataset includes well-known conferences, such as \textit{ICSE}, \textit{FSE}, and \textit{ASE}.
    \item \textbf{SpringerLink.} This dataset contains features journals, such as \textit{Empirical Software Engineering} and \textit{Automated Software Engineering}.
    \item \textbf{ScienceDirect.} This dataset contains relevant journals, such as \textit{Information and Software Technology} and \textit{Journal of Systems and Software}.
    
    \item \textbf{ArXiv.} ArXiv is a freely accessible preprint repository that hosts research papers across various disciplines. It provides an open platform for researchers to share their findings before formal peer review, enabling rapid dissemination of cutting-edge research. Many influential studies in neural code translation first appear on ArXiv, making it a valuable resource for tracking emerging trends and advancements in the field.
    \item \textbf{Google Scholar.} Google Scholar is a widely used academic search engine that indexes a broad range of scholarly literature across disciplines and publication types. Due to its comprehensive coverage, it can supplement the aforementioned academic datasets by retrieving additional studies related to neural code translation that may not be included in those datasets.
\end{itemize}

To retrieve relevant studies on neural code translation from the selected academic databases, we design the following search string: (``code translation" OR ``code-to-code translation" OR ``program translation" OR ``code migration") AND (``deep learning" OR ``large language model"). This search string is intended to capture a broad range of studies that focus on code translation techniques powered by deep learning/large language model approaches.

Finally, we employ the snowballing technique, following prior studies~\cite{liu2022deep,zhang2023survey,wang2024software}, to ensure a comprehensive collection of primary research. Specifically, we examine the reference lists (i.e., backward snowballing) and citations (i.e., forward snowballing) of the initially identified studies to uncover additional relevant work. This technique helps identify studies that may not be directly retrieved through the search string, particularly those that adopt alternative terminology. By applying both backward and forward snowballing, we aim to mitigate the risk of missing important and relevant publications.

\subsection{Study Selection}

To select high-quality primary studies, we first show the inclusion criteria for our SLR on neural code translation.

\begin{itemize}
    \item \textbf{Focus on neural code translation.} The selected study should focus on the topic of neural code translation.    
    \item \textbf{Language of publication.} The selected study should be written in English.    
    \item \textbf{Methodology.} The selected study should either propose a new method or conduct a large-scale empirical study.    
    \item \textbf{Evaluation of results.} The selected study should evaluate the performance of neural code translation approaches.    
\end{itemize}

Then, we show exclusion criteria for our SLR on neural code translation.

\begin{itemize}
    \item \textbf{Non-English papers.}: Studies written in languages other than English should be excluded.    
    \item \textbf{Non-neural code translation focus.} Studies focusing on other similar software engineering tasks (such as code generation and code optimization) should be excluded.    
    \item \textbf{Lack of empirical evaluation.} Studies that lack empirical evaluation should be excluded.    
    \item \textbf{Duplicated publications.} If the study is an extended journal version of a conference paper, the conference version of this study should be excluded.    
    \item \textbf{Limited paper length}: Some studies, such as short papers, NIER (New Ideas and Emerging Results) papers, and tool demo papers, should be excluded due to their limited length and lack of rigorous evaluation.    
\end{itemize}

In a systematic literature review, applying quality assessment criteria is essential to ensure the high quality of the selected studies.
To guarantee the inclusion of high-quality studies in this SLR, we have developed a comprehensive set of quality assessment criteria, following the guidelines designed by Hou et al.~\cite{hou2024large}. These criteria are utilized to systematically evaluate the relevance, clarity, methodological rigor, and overall significance of the candidate papers. The detailed quality assessment criteria are presented in Table~\ref{tab:qa_criteria}.

\begin{table}[ht]
\centering
\caption{Quality assessment criteria for neural code translation studies}
\resizebox{\textwidth}{!}{%
\begin{tabular}{|p{5cm}|p{8cm}|p{3cm}|}
\hline
\textbf{Criterion} & \textbf{Description} & \textbf{Scoring} \\ \hline
QA1: Relevance & Does the study explicitly focus on neural code translation? & Yes/No \\ \hline
QA2: Research objectives & Are the research objectives aligned with neural code translation? & Yes/Partially/No \\ \hline
QA3: Methodology description & Does the study provide a clear methodology description for neural code translation? & Yes/Partially/No \\ \hline
QA4: Model details & Does the study specify the neural model used (e.g., Transformer, CodeT5, ChatGPT) and describe how it was applied? & Yes/Partially/No \\ \hline
QA5: Evaluation metrics & Are the evaluation metrics clearly defined? & Yes/Partially/No \\ \hline
QA6: Evaluation subject description & Does the study describe the subjects for evaluation? & Yes/No \\ \hline
QA7: Practical contribution & Does the study provide practical insights or tools for practitioners in neural code translation? & Yes/Partially/No \\ \hline
QA8: Replicability & Does the study provide sufficient details to allow experiment replication? & Yes/Partially/No \\ \hline
QA9: Novelty and Originality & Does the study propose a novel approach or significant improvement over baselines? & Yes/No \\ \hline
QA10: Publication Type and Venue &  Is the study published in a peer-reviewed journal, conference, or workshop? If published on arXiv, does it satisfy the quality criteria? & Yes/No \\ \hline
\end{tabular}
}
\label{tab:qa_criteria}
\end{table}

Following previous SLRs~\cite{wang2024software,hou2024large,jiang2024survey,zhang2024systematic}, each quality assessment criterion adopts either a binary or ternary scoring scheme, depending on the nature of the question. For binary items (e.g., QA1, QA6, QA9, QA10), scores are assigned as 1 (Yes) and -1 (No) based on the level of fulfillment. For ternary items (e.g., QA2, QA3, QA4, QA5, QA7, QA8), scores are assigned as 1 (Yes), 0 (Partially), and -1 (No). The total score for each study is calculated by summing the scores across all ten criteria, with a maximum possible score of 10.
We set the inclusion threshold at 8 points (80\% of the total score). Only studies that meet or exceed this threshold are included for further analysis.
This quality assessment process can help to systematically exclude studies with misaligned objectives, vague methodologies, or limited contributions, thereby ensuring that the final set of included studies is representative, rigorous, and of high reference value.

As a result, we selected {\primarystudies} primary studies for conducting our SLR. Next, we analyzed the selected studies from two perspectives: the number of publications over different years and the venues where the studies were primarily published.

\begin{figure}[h!] 
    \centering 
    \subfloat[Publication number over different years]{\includegraphics[width=0.45\textwidth]{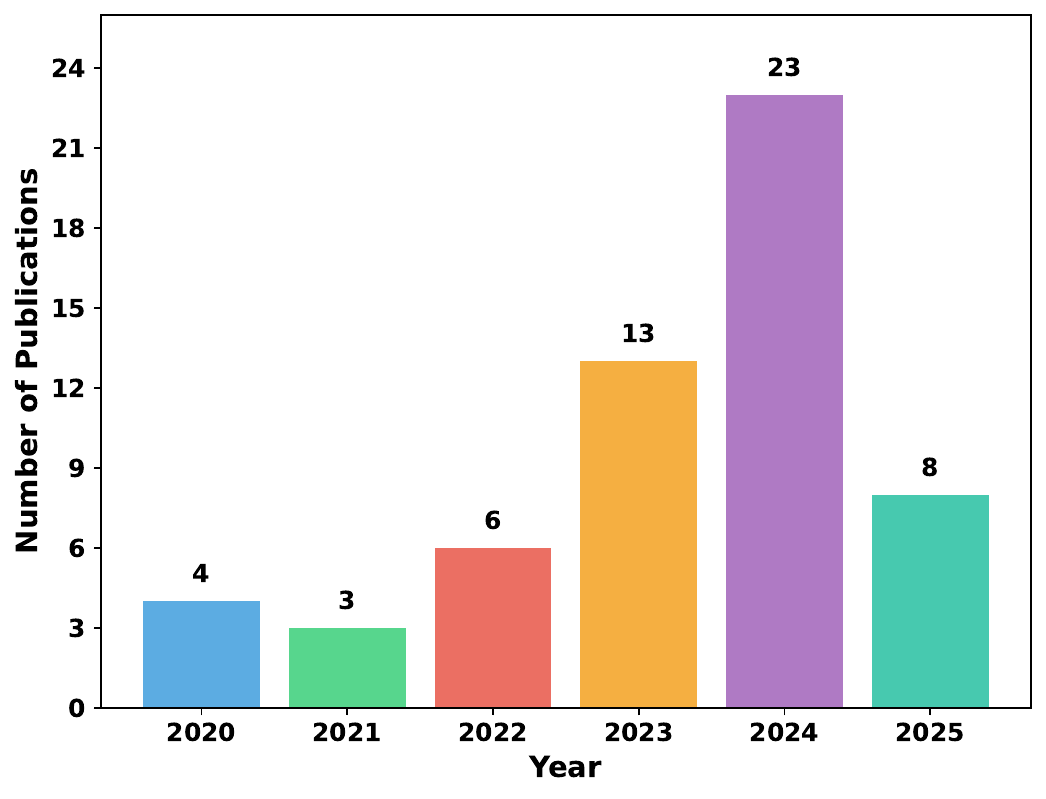} \label{Fig:year}} 
    \hspace{0.5cm} 
    \subfloat[Word cloud of major venues]{\includegraphics[width=0.45\textwidth]{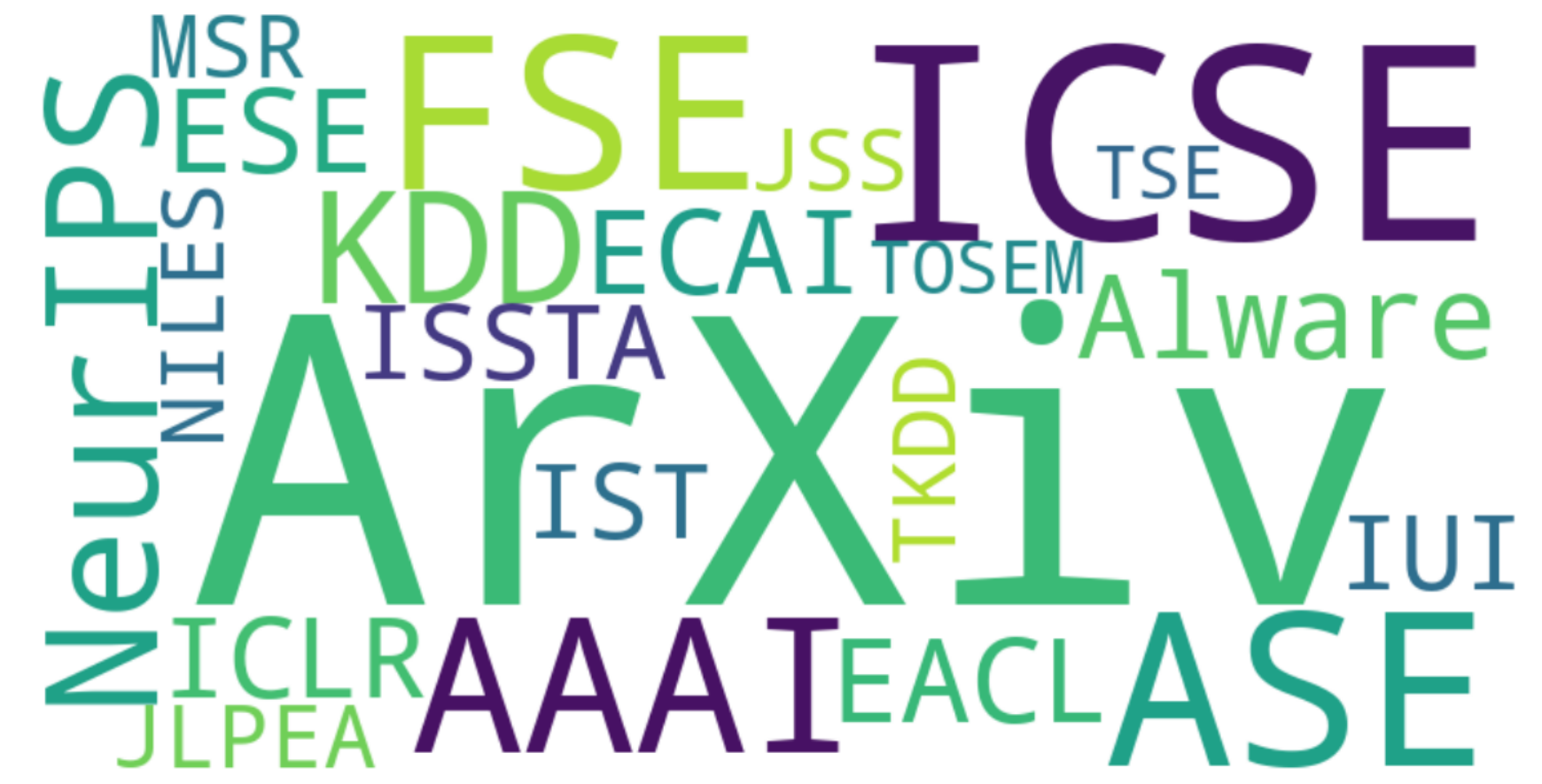} \label{Fig:venue}} 
    \caption{Publication trends and major venues, (a) publication number over different years, (b) word cloud of major venues} 
    \label{Fig:combined}
\end{figure}

\textbf{Analysis on publication year.} Fig.~\ref{Fig:year} presents the number of selected studies over different years. 
The trend of publications on neural code translation shows a steady increase from 2020 to 2024. Specifically, the number of publications remained relatively low from 2020 to 2022, with a slight rise from 4 in 2020 to 6 in 2022. However, a significant surge occurred in 2023, reaching 13 publications, and peaked in 2024 with 23 studies, indicating a growing research interest in this field. The decline in 2025 (8 publications) is attributed to incomplete data collection for the ongoing year, as we only collected relevant papers published before March 2025.

\textbf{Analysis on publication venues.} 
Then, we analyzed the major venues where these studies were published, and Fig.~\ref{Fig:venue} visualizes these publication venues via a word cloud. 
The word cloud illustrates the most frequent publication venues for neural code translation research. Larger and more prominently displayed venues indicate higher publication frequency. Key venues include: (1)
ICSE (International Conference on Software Engineering), FSE (Foundations of Software Engineering), ISSTA (International Symposium on Software Testing and Analysis), and ASE (Automated Software Engineering) are top-tier software engineering conferences.
(2) ArXiv, a popular preprint repository where many early-stage and influential works are shared.
(3) AAAI (Association for the Advancement of Artificial Intelligence), ICLR (International Conference on Learning Representations), and NeurIPS (Conference on Neural Information Processing Systems), highlighting interest from the artificial intelligence and machine learning communities.
(4) TKDD (Transactions on Knowledge Discovery from Data) and KDD (Knowledge Discovery and Data Mining), which emphasize automation in data mining and data-driven methods.
(5) Other venues like TOSEM (ACM Transactions on Software Engineering and Methodology), TSE (Transactions on Software Engineering), ESE (Empirical Software Engineering), IST (International Conference on Imaging Systems and Techniques), and JSS (Journal of Systems and Software) indicate the presence of relevant journal publications.
This distribution suggests that (1) neural code translation research is a highly interdisciplinary field, spanning software engineering, data mining, and machine learning; (2) Most of the studies were published in top-tier conferences and journals; (3) it has become a recent active research topic in AI for software engineering, as evidenced by the large number of related papers published on arXiv.

\subsection{Data Extraction}

Data extraction is a critical phase in the SLR process, during which relevant information is systematically retrieved from the selected studies to address the research questions.
This stage ensures the consistency and completeness of the extracted data, which is essential for synthesizing findings across diverse studies.

For each study, we extract the following key information:

\begin{itemize}
    \item \textbf{Study information.} Author(s), year of publication, title, and venue (such as journal or conference).
    \item \textbf{Research objective.} The main focus or goals of the study related to neural code translation.
    \item \textbf{Neural models used.} The specific models (e.g., Transformer, CodeT5, or ChatGPT) employed in neural code translation.
    \item \textbf{Methodology.} The approach used for neural code translation (e.g., training from scratch, fine-tuning, or prompt engineering).
    \item \textbf{Evaluation metrics.} The metrics used to evaluate the effectiveness of neural code translation approaches (e.g., code similarity, or test case pass rate).
    \item \textbf{Benchmark/Datasets.} The datasets or benchmarks used for evaluating the effectiveness of neural code translation approaches.
    \item \textbf{Key Findings.} Main results and key implications of the study, including effectiveness, challenges, and contributions.
    \item \textbf{Limitations and future directions.} Any limitations mentioned by the authors, as well as suggested future research directions.
\end{itemize}

\subsection{Data Synthesis}

Once the relevant data had been extracted, it was synthesized to identify key trends, research gaps, and common patterns across the selected studies. This information was then analyzed to answer the research questions and provide insights into the current state of research in neural code translation.
To ensure the accuracy and reliability of the extracted data, two reviewers (i.e., the first and third authors) independently carried out the data extraction process. Any discrepancies were resolved through discussion until a consensus was reached.
By following a structured and rigorous data extraction procedure, this SLR ensures the systematic collection of all relevant information, thereby enabling a comprehensive and reliable analysis of the progress and challenges in neural code translation research.

%% file: sections/4.resultanalysis.tex
\section{Result Analysis}
\label{sec:result}

\subsection{RQ1: Task Characteristics}

To analyze the task characteristics of neural code translation, we primarily consider two key aspects: the programming language pairs involved and the granularity of the code.

\subsubsection{Programming Language Pairs}

In programming languages, the type system defines and manages the data types of variables and expressions. 
Type systems are generally categorized into two main types: \textbf{Statically Typed} and \textbf{Dynamically Typed}.
Statically typed languages perform type checking at compile time, with representative examples including Java, C++, and Go. 
In contrast, dynamically typed languages perform type checking at runtime, such as Python, Ruby, and JavaScript.
Based on the classification of type systems, code translation between different programming languages (i.e., programming language pairs) can be classified into four categories:

\begin{itemize}
    \item \textbf{Statically Typed $\rightarrow$ Statically Typed.}
This type of translation involves converting code from one statically typed language to another. A key advantage is that type information is available at compile time, enabling a more direct mapping between the type systems of the source and target languages. However, despite the shared static typing, the translation process must still account for differences in syntax, semantics, and standard libraries. A typical example of this category is translating C++ code to Rust.

    \item \textbf{Statically Typed $\rightarrow$ Dynamically Typed.}
This type of translation converts code from a statically typed language to a dynamically typed one. A key characteristic is the removal or relaxation of explicit type declarations, such as eliminating `int`  annotations. To compensate for the absence of compile-time type checking, runtime type-checking mechanisms may be introduced during translation. A common example is translating C\# code to Python, which often requires explicitly handling type constraints and ensuring semantic consistency.

    \item  \textbf{Dynamically Typed $\rightarrow$ Statically Typed.}
This type of translation involves converting code from a dynamically typed language to a statically typed one. The primary challenge lies in supplementing missing type information, which may require type inference or manual annotations. The translated code may introduce generics or enforce strict type constraints, potentially reducing the flexibility inherent in the original code. A typical example is translating Python code to Go, where developers often need to manually add type annotations or leverage tools like ``mypy`` to infer types before translation.

    \item  \textbf{Dynamically Typed $\rightarrow$ Dynamically Typed.}
This type of translation involves converting code from one dynamically typed language to another. Since both the source and target languages utilize dynamic type systems, explicit type conversions are generally unnecessary. However, the translation process can be complicated by differences in syntax and runtime behavior, such as variations in weak type conversion rules.
Compared to statically typed language pairs, dynamic-to-dynamic translation is typically more straightforward. However, it may introduce hidden type-related errors due to the absence of strict compile-time type enforcement. A representative example is translating JavaScript to Lua, which requires careful handling of runtime differences, including event loop models and asynchronous execution models.

\end{itemize}

Based on the classification above, it is clear that code translation between programming languages with different type systems presents distinct challenges. Statically typed languages enforce type checking at compile time, while dynamically typed languages perform type checking at runtime. This fundamental difference requires careful mapping of types and handling of implicit conversions to avoid potential errors.
Moreover, each programming language has its own syntax and programming paradigms. As a result, translated code must not only be functionally equivalent but also clear, maintainable, and consistent with the best practices of the target language.

\begin{table*}[htbp]
\caption{Task characteristic summary of neural code translation in terms of programming language pairs}
\centering
\resizebox{1.0\textwidth}{!}{
\begin{tabular}{|c|c|c|l|}
\hline
\textbf{Translation Type} & \textbf{Categories of PL Pair} & \textbf{PL Pair} & \textbf{Related Studies} \\
\hline
\multirow{24}{*}{Bidirectional Translation} 
& \multirow{15}{*}{Statically Typed $\leftrightarrow$ Statically Typed} 
& Java $\leftrightarrow$ C\# & \cite{guo2020graphcodebert, wang2022no, zhu2022multilingual, liu2023contrabert, jha2023codeattack, chen2023exploring, huang2023program, li2024few, tipirneni2024structcoder, zhu2024grammart5, agarwal2021using, ahmad2021unified, lin2022xcode, chakraborty2022natgen,kumar2025can,du2024joint} \\
 & & Java $\leftrightarrow$ C++ & \cite{roziere2020unsupervised, agarwal2021using, roziere2021leveraging, jiao2023evaluation, szafraniec2022code, lin2022xcode, zheng2023codegeex, liu2023syntax, liu2024mftcoder, yang2024exploring, di2024codefuse, zhu2024semi, yin2024rectifier, macedo2024intertrans, yuan2024transagent,xue2025classeval,du2025post,pan2024lost,du2024joint} \\
  & &Java $\leftrightarrow$ Go & \cite{szafraniec2022code, zheng2023codegeex, macedo2024intertrans,pan2024lost} \\
   & & Java $\leftrightarrow$ Rust & \cite{szafraniec2022code, macedo2024intertrans} \\
 & & Java $\leftrightarrow$ C & \cite{zhu2024semi,pan2024lost,du2024joint} \\
 & & C++ $\leftrightarrow$ C\# & \cite{lin2022xcode,pan2024lost,du2024joint} \\
  & & C $\leftrightarrow$ C\# & \cite{du2024joint} \\
 & & C++ $\leftrightarrow$ C & \cite{zhu2024semi,du2024joint} \\
  & & C++ $\leftrightarrow$ Rust & \cite{szafraniec2022code, macedo2024intertrans} \\
 & & C++ $\leftrightarrow$ Go & \cite{szafraniec2022code, zheng2023codegeex, macedo2024intertrans,pan2024lost} \\
  & & C $\leftrightarrow$ Go & \cite{pan2024lost} \\
 & & Swift $\leftrightarrow$ Java & \cite{muhammad2020trans} \\
& & C++ $\leftrightarrow$ CUDA& \cite{tehrani2024coderosetta} \\
& & Fortran $\leftrightarrow$ C++ & \cite{tehrani2024coderosetta} \\
& & Rust $\leftrightarrow$ Go & \cite{szafraniec2022code, macedo2024intertrans} \\
\cline{2-4}
& \multirow{3}{*}{Dynamically Typed $\leftrightarrow$ Dynamically Typed} 
&  Python $\leftrightarrow$ JavaScript & \cite{zheng2023codegeex, macedo2024intertrans,du2024joint} \\
&&  PHP $\leftrightarrow$ JavaScript & \cite{du2024joint} \\
&&  PHP $\leftrightarrow$ Python & \cite{du2024joint} \\
\cline{2-4}
& \multirow{15}{*}{Statically Typed $\leftrightarrow$ Dynamically Typed} 
& Python $\leftrightarrow$ C++ & \cite{roziere2020unsupervised, agarwal2021using, roziere2021leveraging, lin2022xcode, zheng2023codegeex, liu2023syntax, liu2024mftcoder, yang2024exploring, di2024codefuse, zhu2024semi, yin2024rectifier, macedo2024intertrans, yuan2024transagent,xue2025classeval,du2025post,pan2024lost,du2024joint} \\
& & Python $\leftrightarrow$ Go & \cite{zheng2023codegeex, macedo2024intertrans,pan2024lost} \\
 & & Python $\leftrightarrow$ Rust & \cite{macedo2024intertrans} \\
 & & JavaScript $\leftrightarrow$ Rust & \cite{macedo2024intertrans} \\

& & Java $\leftrightarrow$ Python & \cite{malyala2023ml, ahmad2023summarize, li2024few, jana2024cotran, roziere2020unsupervised, agarwal2021using, roziere2021leveraging, jiao2023evaluation, lin2022xcode, zheng2023codegeex, liu2023syntax, liu2024mftcoder, yang2024exploring, di2024codefuse, zhu2024semi, yin2024rectifier, macedo2024intertrans, yang2025assessing, yuan2024transagent,liu2024hmcodetrans,xue2025classeval,du2025post,pan2024lost,du2024joint} \\
 & & Python $\leftrightarrow$ C & \cite{zhu2024semi,pan2024lost,du2024joint} \\
 & & C++ $\leftrightarrow$ JavaScript & \cite{zheng2023codegeex, macedo2024intertrans,du2024joint} \\
 & & Java $\leftrightarrow$ JavaScript & \cite{zheng2023codegeex, macedo2024intertrans,jiao2023evaluation,du2024joint} \\
 & & C $\leftrightarrow$ JavaScript & \cite{du2024joint} \\
  & & C\# $\leftrightarrow$ JavaScript & \cite{du2024joint} \\
& & C $\leftrightarrow$ PHP & \cite{du2024joint} \\
& & C\# $\leftrightarrow$ PHP & \cite{du2024joint} \\
& & C\# $\leftrightarrow$ Python & \cite{du2024joint} \\
& & C++ $\leftrightarrow$ PHP & \cite{du2024joint} \\
& & Java $\leftrightarrow$ PHP & \cite{du2024joint} \\
\hline
\multirow{12}{*}{Unidirectional Translation} 
& \multirow{7}{*}{Statically Typed → Statically Typed} 
&  C\# → Java & \cite{liu2023improving, shao2025unigencoder} \\
& & Java → C\# & \cite{li2024extracting, chen2024data} \\
& & Java → C++ & \cite{eniser2024automatically} \\
& & Java → Rust & \cite{ou2025enhancing} \\
& &C → Rust & \cite{hong2025type, shiraishi2024context,ou2025enhancing} \\
& & Fortran → C++ & \cite{bhattarai2024enhancing} \\
& & Solidity → Move & \cite{karanjai2024solmover} \\
\cline{2-4}
& \multirow{3}{*}{Dynamically Typed → Dynamically Typed} 
&  JavaScript → Python & \cite{pan2023stelocoder} \\
& &Python → JavaScript & \cite{wang2023transmap} \\
& &PHP → Python & \cite{pan2023stelocoder} \\
\cline{2-4}
& \multirow{3}{*}{Statically Typed →Dynamically Typed} 
& Java → Python & \cite{weisz2022better, xue2024interpretable, pan2023stelocoder, eniser2024automatically, ibrahimzada2025repository} \\
& & C++ → Python & \cite{pan2023stelocoder} \\
& & C\# → Python & \cite{pan2023stelocoder} \\
\cline{2-4}
& \multirow{2}{*}{Dynamically Typed →Statically Typed} 
& Python → Java & \cite{jin2021algorithm} 
\\
&& Python → Rust & \cite{ou2025enhancing} 
\\
\hline
\end{tabular}
}
\label{sec:plpairs}
\end{table*}

\textbf{Discussion.}
Based on the primary studies we have collected, we present the statistical results in Table~\ref{sec:plpairs}. 
Notice Tang et al.~\cite{tang2023explain} and Lano et al.~\cite{lano2024using} also investigated bidirectional code translation. However, these studies involve a large number of programming language pairs, some of which include uncommon languages. Therefore, they were not included in this table.
Specifically,
The statically typed $\rightarrow$ statically typed category includes 32 programming language (PL) pairs, making it the most common category, with a total of 44 studies.
The statically typed $\rightarrow$ dynamically typed category covers 15 PL pairs, ranking second with 29 studies.
The dynamically typed $\rightarrow$ dynamically typed category is the least common, containing only six PL pairs and five studies.
The dynamically typed $\rightarrow$ statically typed category consists of 15 PL pairs, with 26 studies addressing this type of translation.

From the perspective of code translation direction, 16 studies focused on unidirectional translation, while 41 studies addressed bidirectional translation. Unidirectional translation is typically preferred when the task objectives are clearly defined, and reverse translation is unnecessary. In such cases, efficiency and accuracy are prioritized.
In contrast, bidirectional translation is chosen when language interoperability or bidirectional collaboration is essential. However, it comes with higher design and validation costs. Research on bidirectional translation is more closely aligned to ensure language interoperability, while unidirectional translation is often linked to code migration or compilation techniques.

The evolution of programming language pairs in neural code translation research reflects trends in software development, such as the increasing demand for cross-language interoperability, the adoption of emerging programming paradigms, and the rise of domain-specific languages tailored for specialized applications. Before 2020, studies predominantly focused on Java, C++, and Python, with translation efforts centered around pairs such as Java $\leftrightarrow$ Python, C++ $\leftrightarrow$ Java, and C++ $\leftrightarrow$ Python. This emphasis was driven by the widespread use of these languages in large-scale enterprise software, the need for cross-language integration in complex projects, and the growing role of Python in machine learning and data science.
Between 2020 and 2022, the demand for code translation between languages such as Java and C\# increased significantly. This trend was largely driven by the growing need for enterprise software migration, as many large-scale business applications are developed in Java and C\#. As a result, businesses increasingly sought automated solutions to efficiently migrate codebases while preserving functionality and reducing manual effort. 
During this period, unsupervised models like TransCoder~\cite{roziere2020unsupervised}, as well as methods based on pre-trained models such as CodeBERT~\cite{feng2020codebert} and GraphCodeBERT~\cite{guo2020graphcodebert}, demonstrated strong performance improvements for code translation tasks, especially in scenarios involving enterprise-level systems.
From 2022 to the present, the rise of Rust and Go has spurred increased research into code translation C/C++ $\rightarrow$ Rust and Java $\leftrightarrow$ Go. Rust’s emphasis on memory safety and high performance has made it a preferred choice for migrating legacy C/C++ code, especially in system-level and security-critical applications. At the same time, Go’s growing adoption in cloud computing and distributed systems has fueled interest in Java $\leftrightarrow$ Go translation, driven by the need for lightweight, concurrent services.
In parallel, advancements in blockchain technology have led to research on Solidity $\leftrightarrow$ JavaScript/Python translation, motivated by the need to migrate, simulate, and test smart contracts off-chain. These translations also help integrate smart contracts with existing web applications and machine learning pipelines, improving development efficiency and ecosystem compatibility.

\subsubsection{Code Granularity}

In neural code translation research, code granularity refers to the scope of the input and output units involved in the translation process. Different levels of granularity present distinct technical challenges and are suited to different application scenarios. Existing studies on code granularity primarily focus on translation at various levels, including the statement level, function/method level, class/module level, file/project level, and code snippet level.

\begin{itemize}
    
\item \textbf{Statement level.} 
Statement-level translation focuses on capturing the semantics of individual statements in the source language. This sentence-by-sentence approach facilitates verification by enabling direct comparisons between corresponding statements in the source and target languages, thereby aiding in error detection and supporting unit-level testing. Whether implemented through rule-based methods or machine learning models, statement-level translation is generally easier to design, train, and evaluate due to its simplicity and well-defined structure. It is particularly well-suited for migrating legacy systems to modern languages, where statements act as fundamental translation units. However, a key technical challenge lies in the lack of contextual information. Translating statements in isolation (i.e., without considering their surrounding functions, classes, or modules) can lead to the loss of crucial context, potentially resulting in inaccurate or incomplete translations.

\item  \textbf{Function/Method level.} 
This level of granularity considers individual functions or methods as the minimal processing units. Previous studies have shown that function-level granularity is among the most widely adopted approaches. Compared to files or classes, functions provide a balanced length, making them well-suited for single-batch processing while preserving logical coherence~\cite{ahmad2023summarize}. Furthermore, function/method-level translation is particularly common in cross-language code migration scenarios, as it closely reflects real-world development practices. In many cases, developers focus more on migrating or comprehending entire functions rather than isolated code snippets~\cite{wang2022no}.

\item  \textbf{Class/Module level.}
Class/module-level translation treats entire classes or modules as the fundamental units of translation, rather than translating code line by line or at the function/method level. This granularity is particularly well-suited for object-oriented programming languages and multi-file projects, as it helps preserve the structural integrity and semantic consistency of the original code. A representative example is provided by Shiraishi et al.~\cite{shiraishi2024context}, where C code was translated into Rust by consolidating the source into larger modules. This restructuring not only aligned more naturally with Rust’s modular architecture but also facilitated the inclusion of comprehensive declarations and definitions for functions and types.

\item  \textbf{File/Project level.}
File/project-level code translation treats an entire file or even a complete software project as a single translation unit. This coarse-grained granularity helps ensure global consistency and preserves full contextual information, making it particularly suitable for large-scale system migration. However, this approach also introduces several significant challenges. First, the size of a full project may exceed the context window limitations of large language models. Second, differences in syntax and language features across programming languages—for example, Java supports method overloading while Python does not—further complicate the translation process. Third, maintaining functional equivalence between the translated and original code is critical, yet many existing techniques adopt a ``translate first, verify later" strategy, which can hinder timely error detection. Finally, unit tests at this level often involve multiple method calls and long call chains, making it difficult for translated tests to independently verify the correctness of individual methods~\cite{ibrahimzada2025repository}.

\item \textbf{Code snippet level.}
A code snippet refers to a continuous segment of code without strict syntactic boundaries, such as complete statements or functions. This level of granularity is particularly beneficial in low-resource language scenarios (e.g., PHP), where large-scale parallel corpora are scarce, making effective training more challenging~\cite{zhu2022multilingual,du2024joint}.

\end{itemize}

\begin{figure}[htbp] 
    \centering 
    \includegraphics[width=0.8\textwidth]{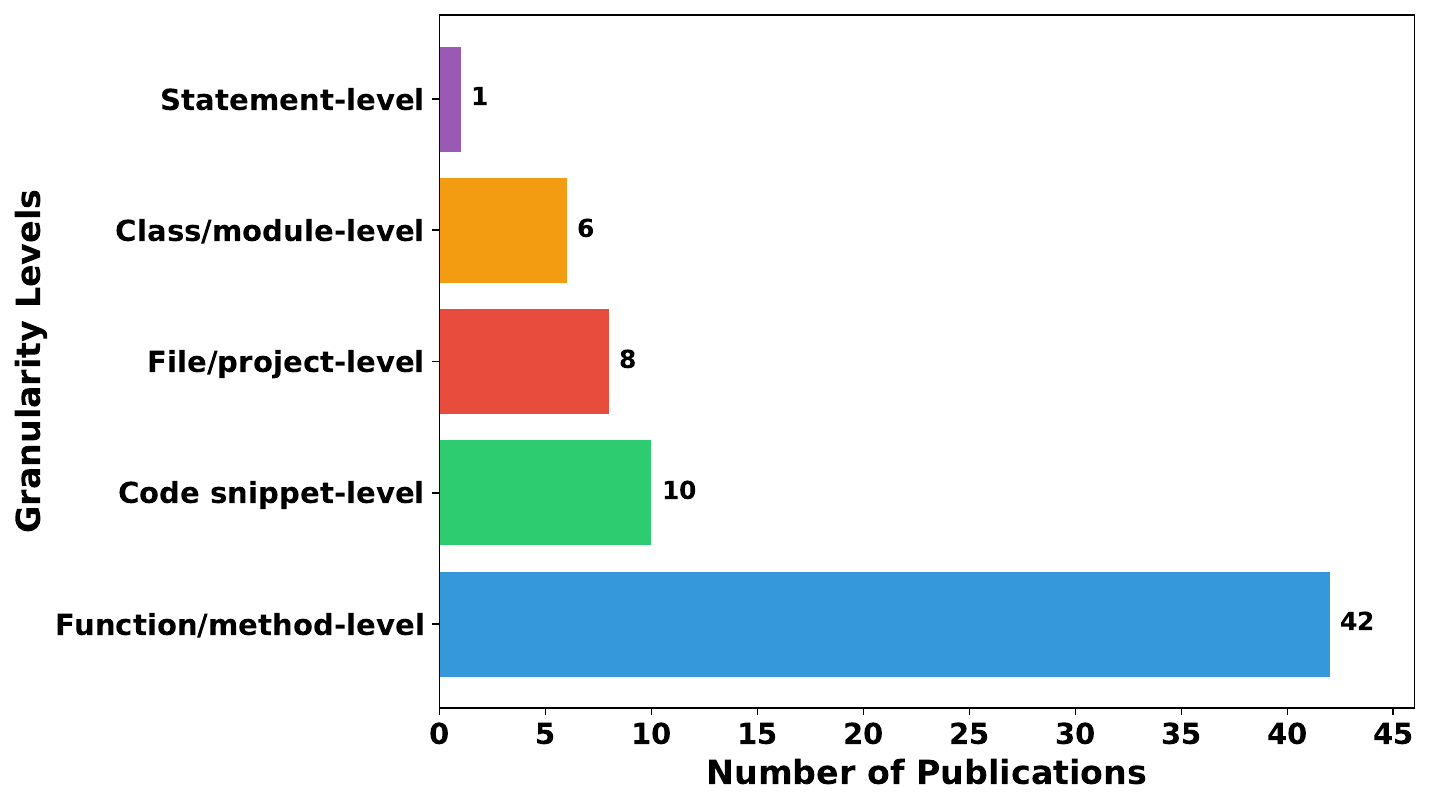} 
    \caption{Task characteristic summary of neural code translation in terms of code granularity} 
    \label{granularity} 
\end{figure}

\textbf{Discussion.}
Fig.~\ref{granularity} presents the statistical distribution of neural code translation studies across different levels of code granularity, ranked by the number of studies. 
Since some studies adopt multiple types of code granularity, the total number of categories exceeds the number of collected papers. For example, Zhu et al.~\cite{zhu2022multilingual} employ both code snippet-level and file/project-level granularity.
From the figure, we observe that function/method-level translation holds the highest proportion, with 42 studies, making it the most extensively explored granularity. The potential reason for this is that functions/methods provide self-contained, meaningful code units that are well-suited for translation while maintaining structural consistency.
In contrast, research at the code snippet-level, file/project-level, and class/module-level is relatively sparse, with 10, 8, and 6 studies, respectively. This may be due to the lack of sufficient contextual information in code snippets, which can result in incomplete semantics, the higher complexity of file/project-level translation in ensuring semantic consistency, and the complexity of class structures and their internal dependencies, which add challenges in ensuring correctness and maintainability during the translation process.
In summary, the trend suggests that previous studies primarily focus on translating smaller, function-level code units. In contrast, code translation at the class level or even project level remains an open challenge in this field due to complex code dependencies, the context window limitations of LLMs, difficulties in testing and validation, and the lack of high-quality benchmark datasets. Addressing these challenges will be essential for advancing neural code translation toward practical, large-scale software migration and modernization.

\begin{tcolorbox}[width=1.0\linewidth, title={}]
\textbf{RQ1: What are the task characteristics of neural code translation?}
\begin{itemize}
    \item In terms of programming language pairs, 71.9\% of studies focus on bidirectional code translation, while 28.1\% focus on unidirectional code translation. The commonly investigated programming language pairs primarily include Java, Python, and C++. Recently, the rise of Rust and Go has made translation targeting these languages a popular research topic.
    \item In terms of code granularity, 73.7\% of studies focus on function/method-level code granularity. However, code translation at the class level or project level would be more practical, but it also requires addressing challenges such as complex code dependencies and the context window limitations of LLMs.
\end{itemize}
\end{tcolorbox}

\subsection{RQ2: Data Preprocessing}

The primary objectives of data preprocessing in neural code translation are to reduce noise, eliminate duplicate data, and perform data augmentation. These steps are essential for enhancing the model’s translation accuracy and improving the quality of the translated code. In the rest of this subsection, we show common data preprocessing methods employed in previous neural code translation studies.

\begin{itemize}
    \item \textbf{Data cleaning.}
Data cleaning is a critical step in improving data quality, aiming to identify and remove erroneous, irrelevant, or low-quality data. Various methods can be employed in this process:
(1) Manual inspection and correction, such as resolving intrinsic logic errors in code or addressing issues related to test code quality. For example, Tang et al.~\cite{tang2023explain} improved the TransCoder evaluation dataset by fixing problems in test cases and correcting errors in both the source and ground truth codes, which led to an average performance improvement of 10\%. Furthermore, other researchers have adopted cross-validation by multiple reviewers to ensure the consistency of manual inspection results~\cite{yang2024exploring,jana2024cotran}.
(2) Rule-based data filtering, which involves identifying and removing low-quality or irrelevant data based on predefined criteria. Common filtering rules include eliminating automatically generated code, abnormal code (e.g., with unusually long average line lengths or low alphabetic character ratios), files that are excessively large or small, and code that fails to meet syntax or quality standards~\cite{zheng2023codegeex,di2024codefuse}.

\item \textbf{Data deduplication.}
Data deduplication is the process of identifying and removing duplicate records from a dataset. Duplicate data can arise from various sources, such as merging datasets from different repositories. Removing these duplicates is essential for ensuring data integrity, preventing model overfitting, and enhancing the generalization ability of neural code translation.
For example, Yang et al.~\cite{yang2025assessing} discovered duplicate code in the AVATAR dataset and addressed this issue by eliminating redundant data. Similarly, Di et al.~\cite{di2024codefuse} improved data quality and diversity through multi-level deduplication, including file-level deduplication using MD5 hashing and code snippet-level deduplication based on program analysis.

\item \textbf{Data augmentation.}
Data augmentation is a technique used to artificially expand and diversify training datasets. In the context of neural code translation, its primary goal is to increase data diversity, reduce the model’s reliance on specific coding styles or variable naming conventions, and ultimately enhance the robustness and accuracy of code translation. Common augmentation strategies include variable and function name replacement while preserving semantic correctness, as well as code style modifications through reformatting to reflect different programming conventions. Another effective approach involves the generation of adversarial examples. For instance, Yang et al.~\cite{yang2025assessing} employed an exhaustive heuristic algorithm to systematically identify and generate the most challenging adversarial examples, thereby improving model robustness through targeted augmentation.

\item \textbf{Data resampling.}
Data resampling is a technique used to optimize model training by adjusting the distribution of training data to better align with task requirements and mitigate the influence of imbalanced or uninformative samples. For example, Di et al.~\cite{di2024codefuse} improved overall training performance by removing low-resource programming languages with a data proportion below 0.1\%, and by downsampling languages such as HTML, CSS, and JSON, which may negatively impact model learning due to their limited relevance or overrepresentation.

\item \textbf{Other.}
Some studies have proposed data preprocessing techniques tailored to specific programming language pairs. 
For instance, in C$\rightarrow$Rust code translation, customized preprocessing methods were designed to reduce compilation errors. These methods include merging declarations and definitions, reordering code elements, and handling macros~\cite{shiraishi2024context}.

\end{itemize}

\begin{figure}[htbp] 
    \centering 
     \includegraphics[width=0.8\textwidth]{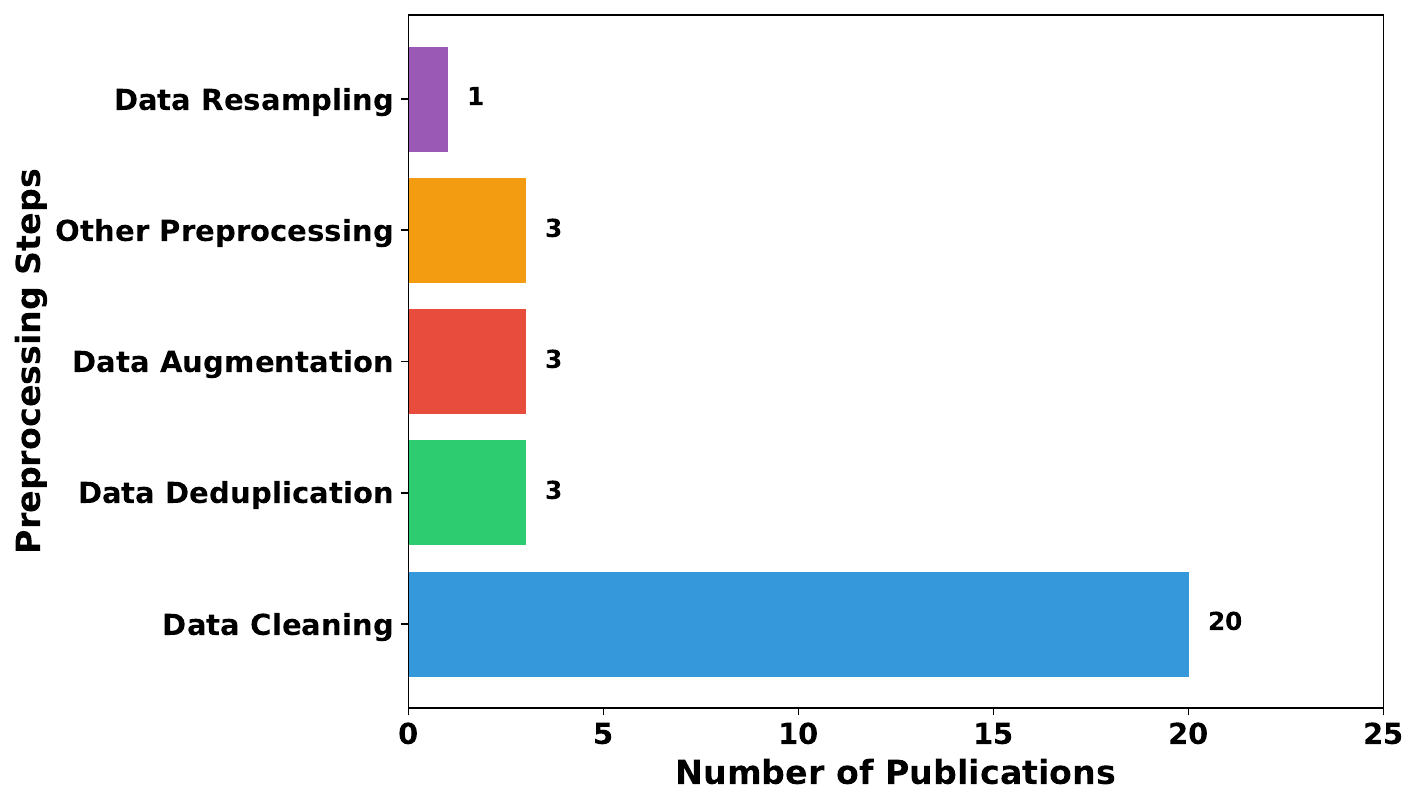} 
    \caption{Data preprocessing summary of neural code translation} 
    \label{preprocessing} 
\end{figure}

\textbf{Discussion.} 
Among the primary studies we collected, 25 studies employed data preprocessing methods. 
Fig.~\ref{preprocessing} illustrates the statistical distribution of neural code translation studies categorized by their use of data preprocessing methods, ranked by the number of associated studies.
Since some studies applied multiple preprocessing techniques, the total number of occurrences across all categories exceeds the total number of collected studies. 
For example, Di et al.~\cite{di2024codefuse} adopted a combination of data cleaning, data deduplication, and data resampling methods.
As shown in the figure, data cleaning emerges as the most widely adopted method, appearing in 20 studies. This highlights its crucial role in enhancing data consistency and reducing noise by removing invalid samples and non-standard code snippets.
Data deduplication is the second most common method, used in three studies, aiming to eliminate redundant samples that could potentially bias model training.
Data augmentation, also applied in three studies, improves dataset diversity by modifying variable or function names and adjusting code formatting, thereby enhancing the model's generalization ability and robustness.
A less commonly employed method is data resampling, utilized in only one study, which addresses data imbalance by adjusting the sample distribution.
Beyond these general strategies, several studies introduced task-specific preprocessing methods tailored to particular programming language pairs. 
In summary, as the saying goes, ``garbage in, garbage out." Data preprocessing is a critical way of improving data quality and ensuring the effectiveness of neural code translation. Among various methods, data cleaning is the most frequently used method. It typically involves manually inspecting code to fix logical errors or designing rules to automatically remove poor-quality code. However, other preprocessing methods have seen limited use. Moving forward, researchers should focus on developing more effective methods for identifying and mitigating data quality issues by following previous studies on code search~\cite{sun2022importance} and vulnerability detection~\cite{croft2023data}. In addition, it is important to design tailored preprocessing strategies that consider the unique characteristics of specific programming languages involved in the translation process.

\begin{tcolorbox}[width=1.0\linewidth, title={}]
\textbf{RQ2: What data preprocessing methods are used for neural code translation?}
\begin{itemize}
    \item In the primary studies we collected, 43.9\% of them utilize data preprocessing methods, with 20 studies specifically focusing on data cleaning.
    \item In the context of neural code translation, researchers primarily rely on traditional data preprocessing methods such as data cleaning and deduplication. Meanwhile, it is noteworthy that some studies are also exploring customized strategies tailored to specific programming language pairs, which deserve further attention in future research.
\end{itemize}
\end{tcolorbox}

\subsection{RQ3: Code Modeling}

In neural code translation, code modeling plays a crucial role in bridging the syntactic, semantic, and structural gaps between programming languages. Its primary objective is to ensure that the translated code remains functionally, logically, and behaviorally consistent with the original. To this end, three major modeling approaches have been widely adopted, each with its strengths and limitations.

\begin{itemize}
    \item \textbf{Treating code as plain text.}
This approach treats source code as plain text and applies sequence-based models (such as Transformers and RNNs) from the field of natural language processing to capture syntax and semantics. Its simplicity allows for effective training on large-scale datasets. However, it often fails to capture the inherent hierarchical structure and logical dependencies in code, which may lead to syntactic errors or loss of structural fidelity. For example, Ahmad et al.~\cite{ahmad2021unified} adopted a sequence-to-sequence model where code snippets were tokenized and processed similarly to natural language text.
    \item \textbf{Modeling code as graphs.}
This method leverages graph-based representations to model both syntactic structure and semantic relationships within code. Commonly used graphs include Abstract Syntax Trees (AST), Control Flow Graphs (CFG), and Data Flow Graphs (DFG), each capturing different aspects such as code hierarchy, execution order, and variable dependencies. These structured representations offer a deeper understanding of code semantics, making them particularly suitable for neural code translation.  
For example, Tipirneni et al.~\cite{tipirneni2024structcoder} proposed StructCoder, in which the encoder processes tokenized source code alongside its AST and DFG using structure-aware self-attention mechanisms.
    \item \textbf{Converting code to intermediate representation (IR).}
IR-based modeling method abstracts the core logic and data flow of code into a language-agnostic form, facilitating semantic consistency and enabling cross-language translation. Despite its potential, this method faces challenges such as conversion complexity, possible loss of language-specific features, and performance overhead.  
For example, Szafraniec et al.~\cite{szafraniec2022code} discussed the difficulties in converting code to IR, particularly in handling constructs that lack direct equivalents across languages.
  \end{itemize}

\begin{figure}[htbp] 
    \centering 
     \includegraphics[width=0.5\textwidth]{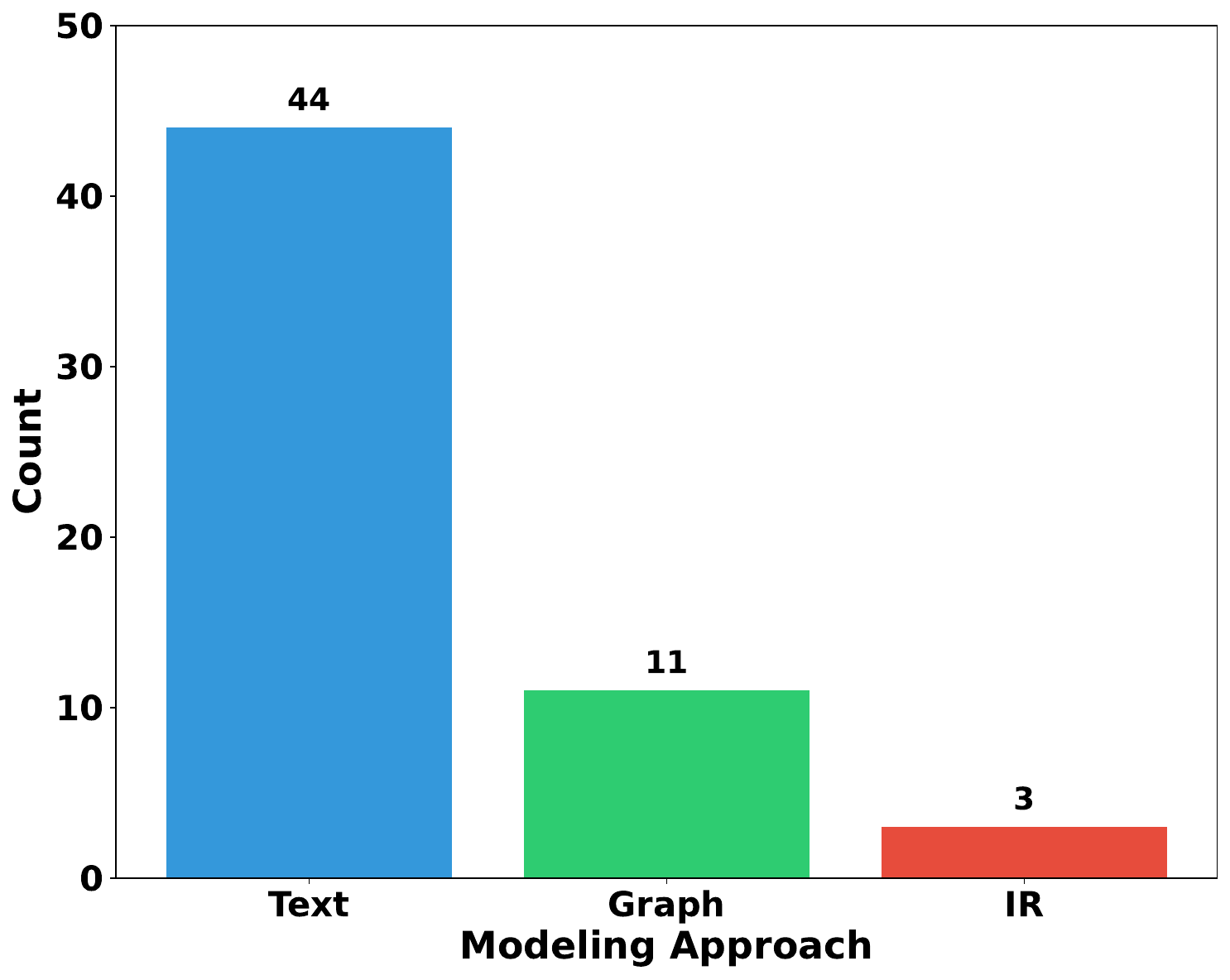} 
    \caption{Code modeling summary of neural code translation} 
    \label{codemodeling} 
\end{figure}

\textbf{Discussion.}
Fig.~\ref{codemodeling} presents the statistical distribution of neural code translation studies based on code modeling methods, ranked by the number of related studies.
Notice that since some studies adopt multiple types of code modeling methods, the sum of all categories exceeds the total number of collected papers.
For instance, Shao et al.~\cite{shao2025unigencoder} proposed the UniGenCoder with two methods: Seq2Seq (sequence-to-sequence) and Seq2Tree (sequence-to-tree). In the Seq2Seq method, code was treated as a regular text sequence, while in the Seq2Tree method, the code was modeled as an AST.
As shown in the figure, treating code as plain text is the most widely adopted modeling method, with 44 studies utilizing this method. Typically, it employs sequence models such as Transformers and RNNs, which can be efficiently trained on large-scale datasets. However, this method does not explicitly model the inherent hierarchical structure, control flow, or data dependencies in code, often resulting in syntactic errors, structural loss, or semantic inconsistencies in the translated output.
Graph-based modeling is the second most common method, appearing in 11 studies. This method uses various graph representations (such as abstract syntax trees, control flow graphs, and data flow graphs) to provide structured representations of code. These representations effectively capture syntactic hierarchy, execution paths, and variable dependencies, thereby improving translation accuracy and semantic preservation. However, graph-based methods also face several limitations. First, constructing such graphs often depends on language-specific parsers and static analysis tools, which can restrict generalizability. Second, the complexity and density of graph structures hinder scalability when dealing with large codebases. Third, integrating graph representations with traditional sequence models remains challenging due to representational differences, limiting their practicality for neural code translation.
Intermediate Representation-based modeling is relatively less explored, with only three studies adopting it. IR abstracts the core logic and data flow of source code in a language-agnostic manner, which helps maintain semantic consistency across languages. However, the IR generation process is complex and may introduce transformation errors or performance overhead. Moreover, language-specific features might be lost during abstraction, which can compromise the quality of the code translation.
In summary, while various code modeling strategies have been investigated in neural code translation, most studies still focus on small-granularity code units (e.g., function or method level), leading to the dominance of plain-text modeling. 
For future research, it is essential to further explore graph-based modeling methods to improve code translation quality and code semantic integrity. In particular, efforts should be made to address existing limitations such as poor scalability and limited cross-language generalization.

\begin{tcolorbox}[width=1.0\linewidth, title={}]
\textbf{RQ3: What code modeling methods are used for neural code translation?}
\begin{itemize}
    \item The code modeling methods in neural code translation mainly include three types: treating code as plain text, modeling code as graphs, and converting code to IR.
    \item In terms of code modeling methods, since the majority of works focus on the code granularity at the method/function level, the most commonly used method is treating code as plain text, which is adopted in 77.2\% of the studies. 
    \item To further improve the quality of code translation, more graph-based modeling methods should be considered, but challenges such as poor scalability and limited cross-language generalization also need to be addressed.
\end{itemize}
\end{tcolorbox}

\subsection{RQ4: Code Translation Model Construction}

The goal of constructing a code translation model is to attempt to train a model that can enable accurate, efficient, and reliable conversion of source code from one programming language to another, while preserving both its functionality and idiomatic style. 
Until now, popular code translation modeling methods can be broadly categorized into five types: training from scratch, fine-tuning, prompt engineering, Agent-based, and RAG-based.

\begin{itemize}
    \item \textbf{Training from scratch.} 
For the code translation task, training from scratch involves building a model entirely without relying on any pre-trained models. The process typically starts with collecting a large-scale parallel code corpus across different programming languages, followed by data cleaning, tokenization, and alignment of source-target code pairs. This is usually done with sequence-to-sequence models (such as the Transformer architecture) to learn the underlying code translation rules from the collected code corpus. While the effectiveness of this approach relies on a large amount of training data and substantial computational resources, it offers flexibility in model design.
For example, Shao et al.~\cite{shao2025unigencoder} first enhanced the Transformer encoder by integrating token sequence generation and strengthened the Transformer decoder by action sequence generation. They then used multi-task learning and knowledge distillation to train the backbone network.

\item \textbf{Fine-tuning.} 
In the context of the code translation task, fine-tuning involves adapting a pre-trained code model (such as CodeBERT or CodeT5) to a specific programming language pair using a relevant code translation corpus. This process typically begins with collecting and preprocessing a parallel code corpus in the source and target languages, which is then used to update the model’s parameters. The fine-tuning method effectively leverages the inherent knowledge of the pre-trained model, allowing it to specialize in the specific code translation task. Compared to training from scratch, fine-tuning can significantly reduce training time and computational resources while maintaining translation quality.
For example, 
Jiao et al.~\cite{jiao2023evaluation} fine-tuned the pre-trained code models CodeT5 and CodeBERT on the XLCoST dataset to perform code translation.
In a recent study, Liu et al.~\cite{liu2024mftcoder} presented a multi-task fine-tuning framework, MFTCoder, based on a popular open-source large language model such as CodeFuse-DeepSeek-33B.

\item \textbf{Prompt engineering.} 
In the code translation task, prompt engineering involves carefully crafting input prompts to guide large language models (such as ChatGPT) in generating the desired code translations. This process typically includes specifying the source and target programming languages, as well as any relevant frameworks or libraries. 
For example, Hong et al.~\cite{hong2025type} guided ChatGPT in code translation using carefully designed prompts, leveraging the model’s training to assist with type migration and function rewriting, thereby ensuring that the translations conformed to the syntax and idiomatic style of the target language.
Du et al.~\cite{du2025post} introduced CAST, which is designed to enhance the performance of LLMs in code translation tasks under the In-Context Learning (ICL) setting by incorporating code structural information to guide exemplar selection.
Ou et al.~\cite{ou2025enhancing} proposed K3Trans, which adopts a Chain-of-Thought (CoT) strategy to guide the LLMs in first identifying the intended functionality of the source function, followed by analyzing the syntactic and dependency differences between the source and target languages.

\item \textbf{Agent-based method.}
Agent-based methods leverage the collaboration of multiple specialized agents, typically built upon LLMs, to systematically handle complex code translation tasks. Each agent in the system is designed to focus on a specific sub-task, such as initial translation, error detection, or refinement, enabling a modular and interactive workflow that improves overall translation quality.
For example, Yuan et al.~\cite{yuan2024transagent} proposed a novel LLM-based multi-agent system, TRANSAGENT, which enhances the effectiveness of LLM-based code translation through the synergy of four LLM agents, including the Initial Code Translator, Syntax Error Fixer, Code Aligner, and Semantic Error Fixer.

\item \textbf{RAG-based method.}
Retrieval-Augmented Generation (RAG) methods integrate retrieval mechanisms into the generation process to enhance code translation performance. By dynamically retrieving semantically relevant code examples or translation pairs from large-scale code repositories and incorporating them into the input context, RAG-based methods enable models to generate higher-quality translated code.
For instance, Bhattarai et al.~\cite{bhattarai2024enhancing} employed a RAG framework that dynamically retrieves semantically relevant code translation examples and strategically incorporates these curated contextual prompts into the model's reasoning process, thereby significantly enhancing the capability of LLMs in the cross-language code translation task.
\end{itemize}

\begin{figure}[htbp] 
    \centering \includegraphics[width=0.6\textwidth]{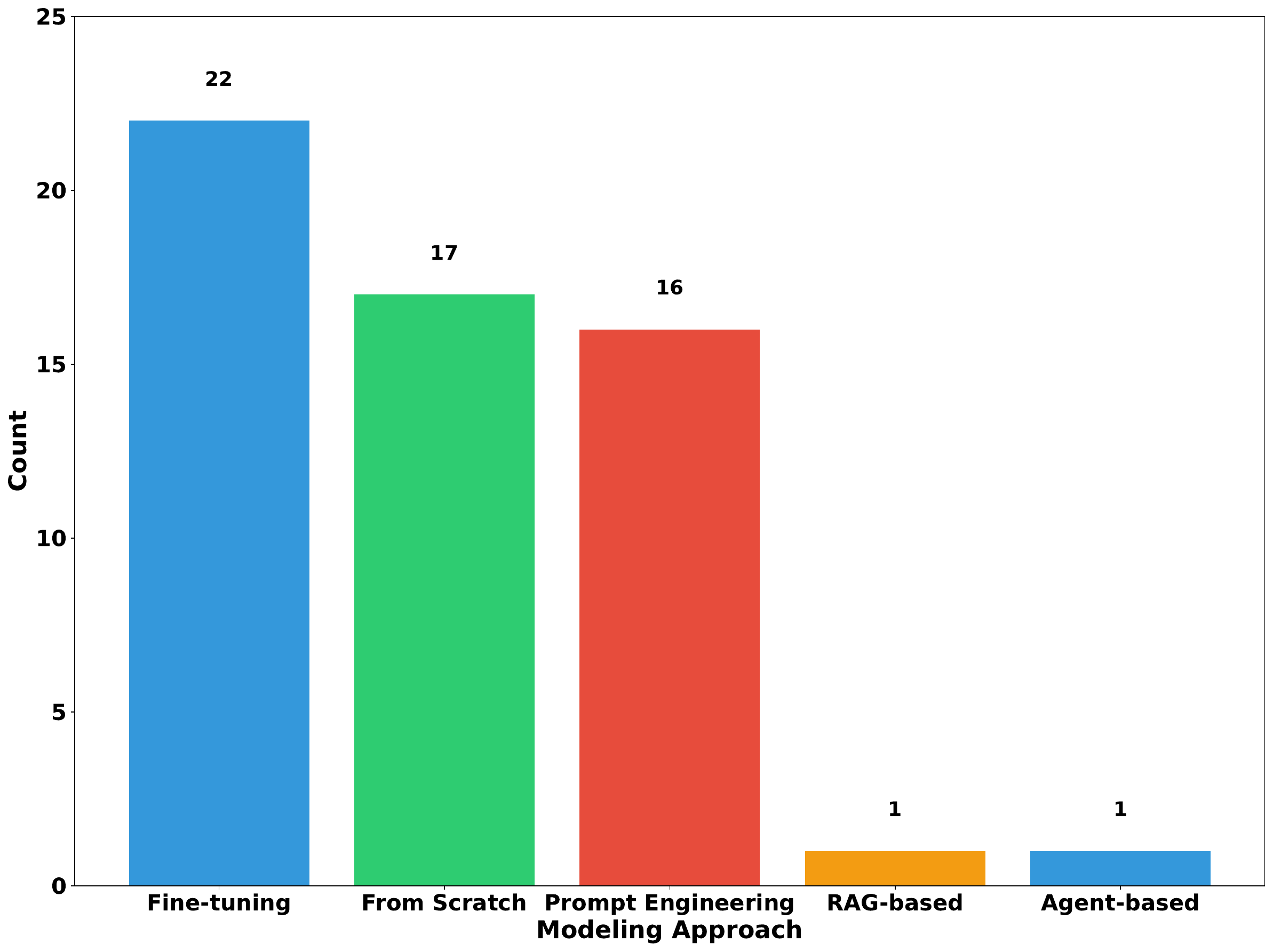} 
    \caption{Code translation model construction summary of neural code translation} 
    \label{construction} 
\end{figure}

\textbf{Discussion.}
Fig.~\ref{construction} illustrates the statistical distribution of neural code translation studies categorized by model construction methods. Among the surveyed studies, fine-tuning emerges as the most widely adopted method, with 22 studies employing this strategy. Its popularity stems from the ability to leverage pre-trained models, thereby reducing the need for extensive task-specific data and computational resources. This method is both efficient and adaptable, enabling models to perform well across a range of code translation tasks.
Training from scratch, used in 17 studies, follows closely. While it allows for the creation of highly specialized models with greater flexibility, it also requires substantial training data and computational resources. Consequently, its broader application is often limited by these high resource demands. Moreover, this approach is primarily adopted by earlier studies, before the widespread availability and adoption of powerful pre-trained models.
Prompt engineering appears in 16 studies, making it the second most popular method. With the increasing scale of large language models, emergent intelligence capabilities have begun to surface, making prompt engineering an emerging and increasingly popular method (particularly effective in zero-shot or few-shot settings). By carefully designing input prompts, this method can guide LLMs (e.g., ChatGPT) to generate accurate translations without modifying their internal parameters. As a result, it offers a cost-effective and practical solution in scenarios with limited resources.
Agent-based methods and RAG-based methods are among the popular research approaches in the recent LLM field. As such, each is currently supported by only one study. It is evident that these methods are still in the early stages of code translation research. However, in the future, more research efforts are expected to explore these directions and propose more effective methods.
In summary, current research trends tend to favor more efficient strategies that leverage pre-trained models, such as fine-tuning or prompt engineering. Although training from scratch remains feasible in highly specialized code translation scenarios, its high resource consumption makes it relatively uncommon in practice.
As research on LLMs continues to advance, integrating emerging techniques such as multi-agent collaboration and RAG is expected to become a key direction for addressing the challenges of complex code translation tasks, offering many new opportunities for future research.

\begin{tcolorbox}[width=1.0\linewidth, title={}]
\textbf{RQ4: What code translation model construction methods are used for neural code translation?} 
\begin{itemize}
    \item Concerning model construction methods, 38.6\% of the studies focus on fine-tuning, making it the most widely adopted approach. Prompt engineering accounts for 29.8\% of the studies, while training from scratch comprises only 28.1\%.
    \item In early research on neural code translation, training from scratch was the dominant approach, despite requiring large-scale corpora and incurring huge computational costs. Recently, more studies have shifted toward fine-tuning and prompt engineering due to the increasing availability of powerful pre-trained models and LLMs.
    \item With the continuous advancement of LLMs, it is essential to consider incorporating emerging techniques (such as multi-agent collaboration and retrieval-augmented generation) to further enhance the performance of the code translation task.
    
\end{itemize}
\end{tcolorbox}

\subsection{RQ5: Post-processing}
Post-processing techniques play a crucial role in improving the quality and accuracy of neural code translation. These methods are designed to address errors, enhance code readability, and ensure alignment with the conventions and requirements of the target language.

\begin{itemize}

\item \textbf{Dynamic-based Bug Detection.}
Dynamic bug detection focuses on identifying semantic inconsistencies and functional errors in translated code by executing the code and analyzing its runtime behavior. 
Unlike static analysis, which inspects code without running it, dynamic-based methods validate translation correctness through test execution and compiler feedback. 
For example, Rozière et al.~\cite{roziere2021leveraging} utilized high-coverage test cases generated by EvoSuite to ensure that translated code preserves the original semantics, filtering out translated codes that fail at runtime. 
Meanwhile, Szafraniec et al.~\cite{szafraniec2022code} analyzed compiler-generated error messages during the execution of translated Rust code to detect semantic violations such as type mismatches or unresolved references.

\item \textbf{Static-based Bug Detection.}
Static-based Bug Detection is an essential method for identifying and addressing bugs in code by using static analysis tools and manual inspections.
For example, Xue et al.~\cite{xue2025classeval} employed a manual inspection approach, involving domain experts to review a portion of their dataset. Through iterative discussions and consensus-building, they developed a comprehensive annotation guideline to ensure consistent and reliable error classification.
Lin et al.~\cite{lin2022xcode} used Tree-Sitter, a parser that analyzes the syntactic structure of the code, to automatically identify grammatical violations in generated code.
Eniser et al.~\cite{eniser2024automatically} employed a static verification technique within the NOMOS framework, which formalized constraints like syntax validity, semantic preservation, and compilability.

\item \textbf{Bug Repair.}
Bug repair in code translation has become a critical step in ensuring the semantic correctness and functionality of translated code. A variety of strategies have been proposed to address different types of errors that may arise during or after translation, including compilation errors, runtime issues, and semantic inconsistencies. Many of these approaches incorporate iterative feedback loops that guide LLMs to refine the translated code based on compiler diagnostics or observed runtime behaviors.
For instance, methods such as TransMap~\cite{wang2023transmap}, INTERTRANS~\cite{macedo2024intertrans}, and AlphaTrans~\cite{ibrahimzada2025repository} adopt line-level mapping, multi-path candidate generation, and feedback-driven refinement techniques, respectively, to detect and correct errors effectively. Compiler error messages are frequently leveraged to guide the repair process in a cost-efficient manner, as exemplified by K3Trans~\cite{ou2025enhancing}, SolMover~\cite{karanjai2024solmover}, and the compiler-based iterative strategies proposed by Hong et al.~\cite{hong2025type}.
To further enhance repair efficiency, some studies restrict the scope of edits using token-level correction methods~\cite{liu2024hmcodetrans} or syntax/semantic alignment techniques~\cite{yuan2024transagent}. Retrieval-based techniques have also been employed to guide editing operations more precisely~\cite{xue2024interpretable}. Additionally, reinforcement learning has been explored to fine-tune translation models based on symbolic or functional verification results~\cite{jana2024cotran}.
Moreover, recent advancements incorporate techniques such as CoT-inspired iterative prompting~\cite{pan2024lost}, memory-based context management~\cite{shiraishi2024context}, and error-guided prompt construction~\cite{yang2024exploring}, which significantly improve the adaptability and flexibility of the repair process.

\item \textbf{Code Optimization.}
Code optimization aims to enhance code quality by improving its maintainability and readability. Common techniques include code refactoring, which involves restructuring existing code to simplify logic, eliminate redundancy, and improve overall clarity without altering its external behavior.
For example, Tang et al.~\cite{tang2023explain} introduced a series of optimization methods that significantly refined code structure and enhanced overall code quality, demonstrating the effectiveness of these strategies.

\end{itemize}

\textbf{Discussion.} 
Among the primary studies reviewed, 23 employed post-processing techniques. Table~\ref{postprocessing} summarizes the categories of post-processing methods used in these studies along with the corresponding references.
The reviewed studies reveal a diverse set of strategies aimed at ensuring the quality of translated code. Dynamic bug detection, which focuses on identifying semantic inconsistencies and functional errors through runtime analysis, was adopted in two studies. By validating the behavior of translated code in real-world execution environments, these methods can uncover issues that static analysis may ignore.
In contrast, static bug detection methods analyze the code without executing it, focusing on structural, syntactic, and semantic correctness. These techniques leverage static analysis tools to detect potential problems early, ensuring that the translated code adheres to expected coding rules. This kind of method was employed in six of our reviewed studies.
Bug repair emerged as another critical area, with 14 studies incorporating iterative feedback loops to refine translated code. These methods commonly use compiler diagnostics, runtime feedback, and error messages to detect and resolve issues, thereby improving both correctness and functionality. Feedback-driven repair strategies play a vital role in enhancing the reliability of the translation process.
Code optimization was adopted in only one study, primarily targeting improvements in the structure, readability, maintainability, and performance of the translated code. In large-scale projects, code comprehensibility and maintainability are critical aspects, as they directly impact the ease of debugging, future development, and long-term sustainability of the software. These factors cannot be overlooked and deserve greater attention in future research.

\begin{table}[htbp]
\centering
\caption{Post-processing categories and related studies}
\resizebox{0.7\textwidth}{!}{
\begin{tabular}{ll}
\toprule
\textbf{Post-processing Category} & \textbf{Related Studies}  \\
\midrule
Dynamic Bug Detection &  ~\cite{roziere2021leveraging,szafraniec2022code}  \\
Static Bug Detection&~\cite{lin2022xcode,eniser2024automatically,zhu2024semi,xue2025classeval,malyala2023ml,jin2021algorithm} \\
Bug Fixing  &~\cite{ou2025enhancing,liu2024hmcodetrans,macedo2024intertrans,karanjai2024solmover,wang2023transmap,jana2024cotran,xue2024interpretable,shiraishi2024context,ibrahimzada2025repository,yin2024rectifier,yang2024exploring,yuan2024transagent,hong2025type,pan2024lost}  \\
Code Optimization   &~\cite{tang2023explain} \\
\bottomrule
\end{tabular}
}
\label{postprocessing}
\end{table}

\begin{tcolorbox}[width=1.0\linewidth, title={}]
\textbf{RQ5: What post-processing methods are used for neural code translation?}
\begin{itemize}
    \item Post-processing techniques are essential in neural code translation for improving output correctness, readability, and alignment with the target programming language’s conventions.
    \item In our collected studies, 40.4\% of them considered post-processing. These operations can be categorized into four types, among which bug repair is the most commonly used.
    \item For code translation in large-scale projects, code comprehensibility and maintainability become even more important and warrant further attention in future research.
\end{itemize}
\end{tcolorbox}

\subsection{RQ6: Evaluation Subjects}

In this section, we introduce the commonly used evaluation subjects for the neural code translation task, presented in chronological order of their release.

\begin{itemize}
    \item \textbf{Chen's dataset.} 
This subject was introduced by Chen et al.~\cite{chen2018tree} to support research on Java–C\# code migration, using real-world open-source projects such as Lucene\footnote{\url{http://lucene.apache.org/}}, POI\footnote{\url{http://poi.apache.org/}}, and JGit\footnote{\url{https://github.com/eclipse/jgit/}}. It consists of 34,628 function-level code pairs that were automatically aligned based on method name and file path similarity, eliminating the need for manual alignment. Ten-fold cross-validation was performed within each project, employing a pairing strategy similar to mppSMT~\cite{nguyen2015divide}. The dataset captures practical migration challenges such as library differences, inconsistent naming, and structural changes, making it a valuable resource for migration rule mining. It is more representative and challenging compared to synthetic datasets, reflecting real-world complexities in code migration.

    \item \textbf{TransCoder.}
This dataset, constructed by Roziere et al.~\cite{roziere2020unsupervised}, collects data from the GeeksforGeeks platform\footnote{\url{https://www.geeksforgeeks.org/}} and over 2.8 million open-source code repositories hosted on Google BigQuery\footnote{\url{https://console.cloud.google.com/marketplace/details/github/github-repos}}. It covers three programming languages (i.e., C++, Java, and Python), with approximately 14,000 function-level samples for each language. The functions, selected from topics like ``graph theory" and ``strings", are short, self-contained, and compilable. During its construction, balanced sampling based on topic tags was applied, and duplicate or excessively long code solutions were filtered out.

    \item \textbf{CodeXGLUE.}
    This subject was introduced by Lu et al.~\cite{lu2021codexglue}. For code translation, they constructed a high-quality Java–C\# parallel corpus for code translation, which was collected from several popular open-source projects hosted on GitHub, including Lucene, POI, JGit, and Antlr\footnote{\url{https://github.com/antlr/}}.
    These projects, originally developed in Java and later migrated to C\#, resulted in 11,800 aligned function pairs. The dataset focuses on Java–C\# translation at the function/method level, enabling fine-grained learning and evaluation.
    To ensure the quality and representativeness of the dataset, several preprocessing steps were implemented. These included mining function pairs based on similar directory structures and function signatures, removing duplicates, and filtering out function pairs with less than one-third token overlap.
    
    \item \textbf{CodeNet.}
    The dataset was released by Puri et al.~\cite{puri2021codenet}. Data was collected from AIZU Online Judge via API access and from AtCoder through web scraping. The dataset covers user-submitted solutions to various algorithmic problems. CodeNet supports 55 programming languages, facilitating multilingual code translation. Over 95\% of the 13.9 million code submissions are written in C++, Python, Java, C, Ruby, and C\#, covering 4,053 problems and totaling approximately 500 million lines of code. Each sample is a single code file that follows a standard input-output format and typically contains a main function. To improve the dataset's quality, deduplication was performed using Jaccard similarity at the file level, and low-quality or erroneous code was filtered out.

    \item \textbf{Avatar.}
    This subject was developed by Ahmad et al.~\cite{ahmad2023avatar}, as a high-quality parallel dataset for Java-Python code translation. It is built from accepted solutions to programming problems sourced from platforms such as AtCoder, AIZU Online Judge, Google Code Jam, Codeforces, GeeksforGeeks, LeetCode, and Project Euler, including both crawled data and GitHub repositories.
    AVATAR covers the Java$\leftrightarrow$Python pair and contains 9,515 problem statements with up to 25 function-level parallel examples per problem. 
    To ensure quality and diversity, they apply tokenization, remove comments and docstrings, filter out long samples (i.e., those exceeding 464 tokens), and select the 5 most diverse solutions for each problem using `difflib'.

    \item \textbf{Other.}  
    Several studies have focused on constructing and optimizing evaluation subjects with the aim of improving their quality, building corpora for low-resource programming languages, and addressing challenges such as data leakage. 
To enhance the quality of evaluation datasets, researchers have developed new resources to ensure the reliability and accuracy of neural code translation. For example, Jana et al.~\cite{jana2024cotran} introduced AVATAR-TC, a large-scale, high-quality cross-language code translation dataset that ensures comprehensive code compilability, syntactic correctness, and functional validation. 
To mitigate the scarcity of parallel corpora for certain programming language pairs, some studies have created specialized benchmarks. For instance, Karanjai et al.\cite{karanjai2024solmover} focused on translating Solidity smart contracts, while Tao et al.\cite{tao2024unraveling} proposed PolyHumanEval, an extension of the HumanEval benchmark to 14 diverse programming languages, including Kotlin, TypeScript, and Rust.
In addition, several efforts have addressed limitations in existing benchmarks, such as data leakage, limited language coverage, and low semantic fidelity. For example, Yuan et al.~\cite{yuan2024transagent} introduced a new high-quality benchmark to overcome these issues.

\end{itemize}

\begin{figure}[htbp] 
    \centering     
    \includegraphics[width=0.6\textwidth]{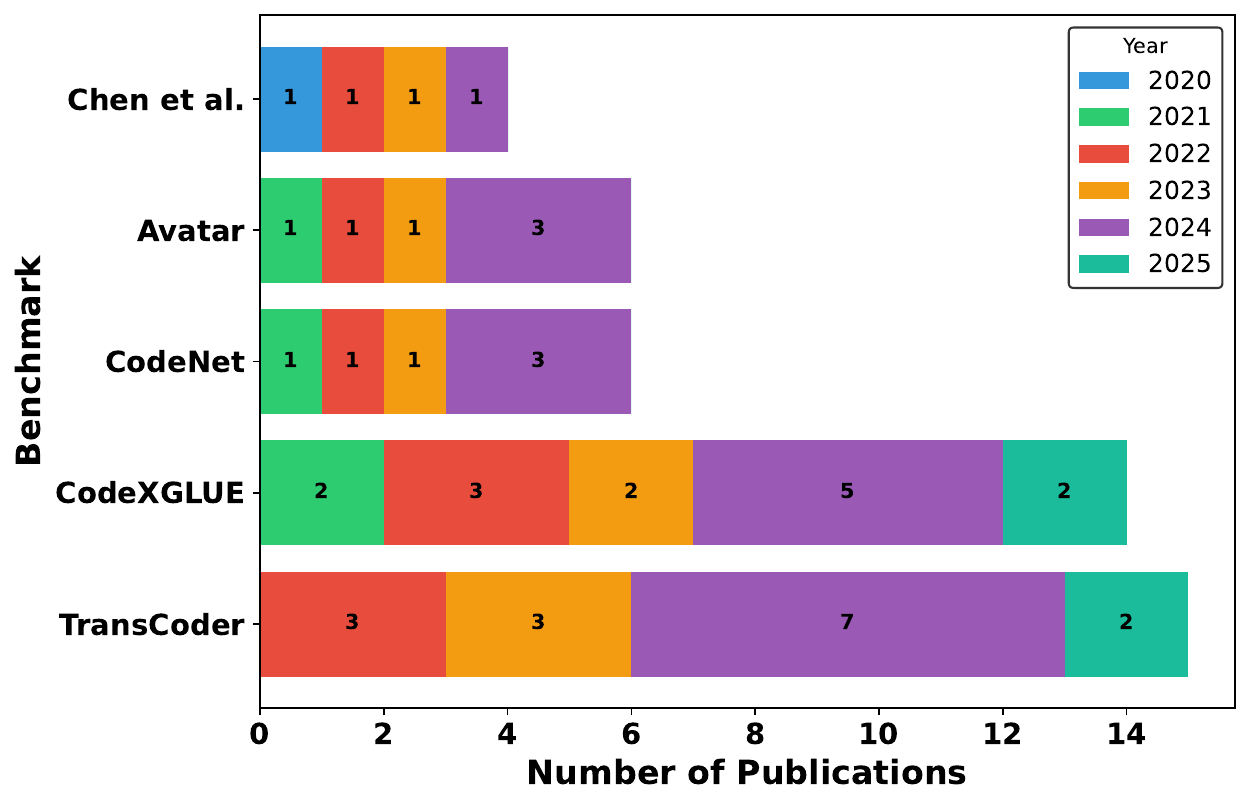} 
    \caption{Evaluation subject summary of neural code translation} 
    \label{subject} 
\end{figure}

\textbf{Discussion.}
Fig.~\ref{subject} illustrates the distribution of neural code translation studies based on commonly used evaluation datasets.
Note that some studies utilize multiple evaluation datasets, and our statistics are based on the evaluation subjects frequently adopted in code translation research; thus, the total count across categories exceeds the number of collected papers.
TransCoder is the most widely used dataset, appearing in 15 studies. It is well-regarded for its unsupervised training approach and large-scale data collected from GeeksforGeeks, making it a strong candidate for general-purpose code translation tasks.
CodeXGLUE is featured in 14 studies and is valued for its comprehensive benchmark suite and support for multiple programming languages. Its versatility and well-structured tasks have led to widespread adoption in the community.
CodeNet and Avatar each appear in six studies. CodeNet is recognized for its massive scale, encompassing 55 programming languages, though it includes a portion of incorrect solutions. In contrast, Avatar is tailored specifically to Java-to-Python translation, providing focused and high-quality samples.
Chen’s Dataset is used in four studies and remains a foundational resource for research on real-world Java–C\# code migration.
Recent advancements in code translation research have increasingly emphasized several key areas.
First, high-quality test cases are essential for validating the functional correctness of translated code. Therefore, integrating comprehensive test suites can significantly improve the reliability of code translation models.
Second, addressing the scarcity of training data for low-resource programming languages, such as Julia and R, has become increasingly important. Specifically, translating code from high-resource languages into low-resource ones, supplemented by synthesized test cases to validate the translations, offers a promising solution.
Third, data leakage poses a significant threat to the integrity of model evaluations, potentially leading to inflated performance metrics. To mitigate this, one approach is to generate perturbed versions of the original code that preserve its functionality.
Finally, understanding code within its broader repository context is crucial for achieving accurate translation. Incorporating repository-level information  (such as the dependency relationships between functions, module structures, and historical code evolution)  can provide valuable context for guiding translation. 
Overall, these recent research directions signify a shift toward more holistic and context-aware solutions in the field of neural code translation.

\begin{tcolorbox}[width=1.0\linewidth, title={}] 
\textbf{RQ6: What are the evaluation subjects for neural code translation tasks?} 
\begin{itemize} 
\item In terms of evaluation subjects, TransCoder is the most frequently studied subject, appearing in 26.3\% of all studies. CodeXGLUE closely follows, representing 24.6\% of the total.
\item In neural code translation, recent subject construction efforts have increasingly emphasized the quality of accompanying test cases, the mitigation of the data leakage issue, and the collection of richer contextual information.
\end{itemize} 
\end{tcolorbox}

\subsection{RQ7: Evaluation Metrics}
Evaluating the quality of neural code translation remains challenging, as it requires ensuring both syntactic correctness and semantic equivalence. To address these challenges, researchers have proposed various evaluation metrics, which can be categorized into two types: static metrics and dynamic metrics.

\textbf{Static-based metrics.} 
Static-based metrics evaluate the translated code by comparing it to ground-truth code using structural or textual similarity, without requiring code execution. Therefore, they offer advantages such as fast evaluation speed and simple implementation. Then we summarize the commonly used static-based metrics as follows:

\begin{itemize}
    \item \textbf{Exact Match.} 
    Exact Match (EM)~\cite{nguyen2013lexical} is a metric used to evaluate the quality of neural code translation by determining whether the translated code exactly matches the ground-truth code. An exact match occurs when every element (such as tokens, syntax, and structure) of the generated code is identical to the ground-truth code without any discrepancies. As a strict metric for accuracy, EM provides a clear and binary measure of translation correctness.
Token Accuracy~\cite{chen2024data, muhammad2020trans}, an extension of EM, measures the percentage of tokens in the translated code that exactly match the corresponding tokens in the ground-truth code. Unlike EM, which requires a complete match of the entire code, Token Accuracy supports partial matches at the token level, making it a more flexible and fine-grained metric for assessing translation quality.

    \item \textbf{BLEU.} 
BLEU~\cite{papineni2002bleu} is a widely used metric for evaluating machine-generated translations and is also applicable to neural code translation. It is based on $n$-gram precision, which measures the overlap between $n$-grams in the generated code and those in the ground-truth code. The BLEU score is typically computed using 1-gram to 4-gram precision, which are then combined using a weighted average. To discourage overly short translations, BLEU incorporates a brevity penalty that penalizes outputs shorter than the reference.
Metrics similar to BLEU include ROUGE~\cite{tehrani2024coderosetta}, METEOR~\cite{lin2022xcode, kumar2025can}, ChrF~\cite{tehrani2024coderosetta}, and CIDEr~\cite{lin2022xcode}. These metrics extend the concept of surface-level $n$-gram matching by incorporating aspects such as semantic similarity, linguistic variation, and recall. As a result, they provide a more comprehensive and nuanced evaluation of translation quality beyond exact token-level overlap.

    \item \textbf{CodeBLEU.} 
CodeBLEU~\cite{ren2020codebleu} is an evaluation metric specifically designed to assess the quality of generated code, addressing the limitations of traditional metrics such as BLEU when applied to programming languages. While BLEU focuses solely on $n$-gram precision, CodeBLEU incorporates additional dimensions (including weighted $n$-gram matching, abstract syntax tree structural similarity, and data-flow consistency) to better capture the syntactic and semantic correctness of code.
Several metrics similar to CodeBLEU have been proposed, including
 $Match_{ast}$~\cite{liu2024hmcodetrans, ou2025enhancing}, $Match_{df}$~\cite{liu2024hmcodetrans}, Dataflow Match~\cite{chakraborty2022natgen}, Syntax Match~\cite{chakraborty2022natgen}, $CodeBLEU_{q}$~\cite{jha2023codeattack}, and DEP~\cite{xue2025classeval}. 
These metrics also incorporate structural and semantic code features, such as abstract syntax trees and data-flow information, to more accurately evaluate the quality of generated code.
   
\end{itemize}

\textbf{Dynamic-based metrics.} 
Dynamic-based metrics evaluate the quality of neural code translation by executing the corresponding test cases. These metrics assess the correctness of the translated code based on the results of test case execution. As such, they offer advantages in measuring the practical correctness of the code translation. However, they often require high-quality test cases, longer evaluation times, and an appropriate execution environment. Then, we summarize the commonly used dynamic-based metrics as follows:

\begin{itemize}
    \item \textbf{Computational Accuracy.} 
       Computational Accuracy (CA)~\cite{roziere2020unsupervised} measures the proportion of translated codes that return the same execution results as the ground truth codes when given the same test cases.
      Several metrics belong to the CA category. For example, 
    FEqAcc~\cite{jana2024cotran} denotes the proportion of translations that achieve functional equivalence under manually designed test cases. 
    DSR@k~\cite{ou2025enhancing} evaluates whether incorrect translations can be debugged to pass all test cases within $k$ attempts, reflecting the model’s debuggability.

    \item \textbf{Pass@$k$.}  
Pass@$k$~\cite{chen2021evaluating} measures the proportion of code translation cases in which at least one of the $k$ generated translated code samples can successfully pass the test cases.
    
    \item \textbf{Compilation Success Rate.}
    Compilation Success Rate (CSR)\cite{xue2025classeval} measures the proportion of translated code samples that can successfully compile.
    In contrast, Compilation@$k$\cite{ou2025enhancing} measures the proportion of code translation cases in which at least one of the $k$ generated translated code samples can successfully compile.

\end{itemize}

\textbf{Other.}
In addition to similarity-based static metrics and execution-based dynamic metrics, other types of evaluation metrics have been proposed. Some researchers leverage static analysis tools or parsers to detect syntax and type errors, enabling the assessment of whether the translated code is usable without the need for execution. Representative metrics in this category include Code-Exec~\cite{yang2025assessing} and Error Rate~\cite{lin2022xcode}.
Other researchers focus on evaluating the human effort required for code verification and repair. For instance, Liu et al.~\cite{liu2024hmcodetrans} introduced Word Stroke Ratio, Key Stroke Ratio, and Mouse Action Ratio, which respectively measure the ratio of modified words, the number of keystrokes required for character changes, and the number of mouse actions per character.

\begin{figure}[htbp]
    \centering 
     \includegraphics[width=0.8\textwidth]{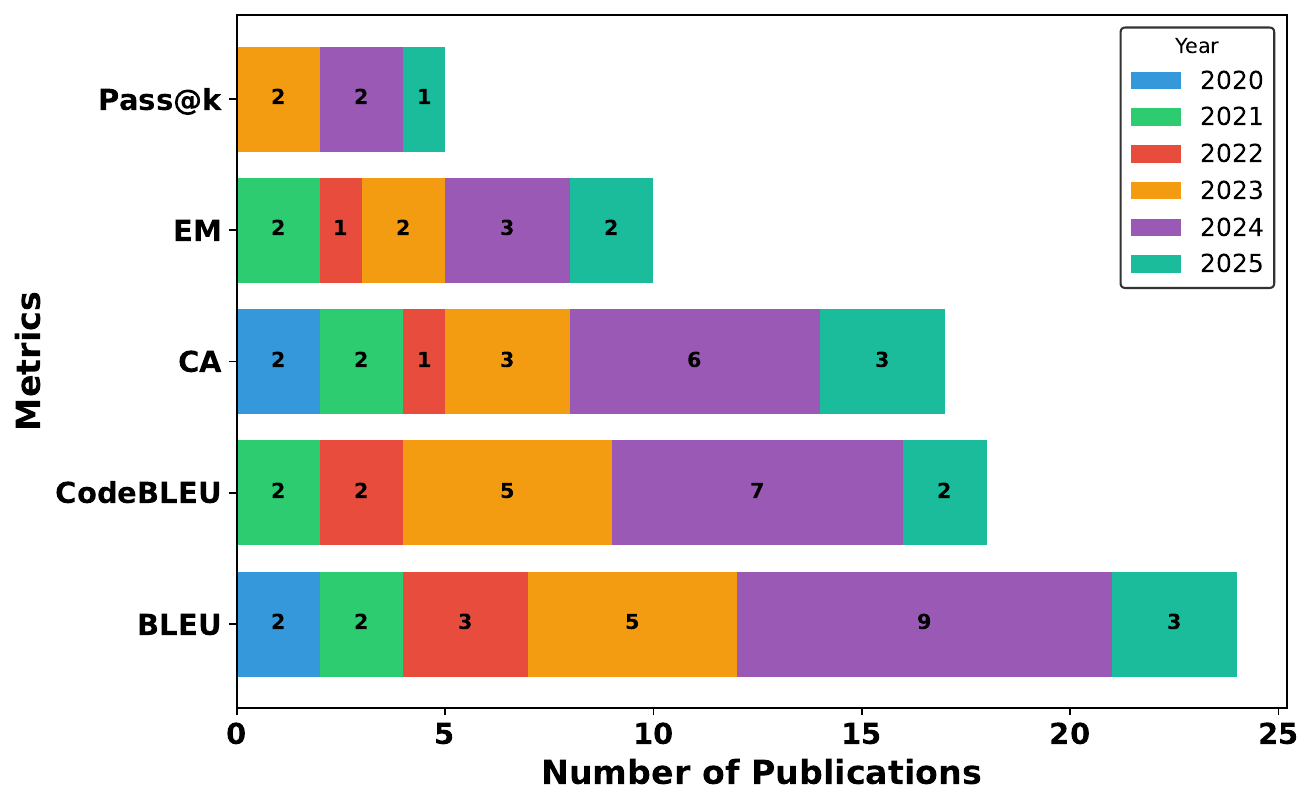} 
    \caption{Top-5 evaluation metrics used for neural code translation} 
    \label{metrics} 
\end{figure}

\textbf{Discussion.} 
Fig.~\ref{metrics} presents the top five evaluation metrics employed in neural code translation studies. Static metrics such as BLEU and CodeBLEU are the most widely adopted, appearing in 24 and 18 studies, respectively. Their popularity underscores the significance of measuring textual similarity as well as the syntactic and semantic correctness of translated code. Exact Match is also frequently used, with 10 studies leveraging it to assess strict correctness by requiring the generated code to match the ground truth exactly.
However, static metrics can be misleading in code translation tasks, as models may achieve high scores while producing code that fails to compile or execute correctly~\cite{pan2024lost}. To address this limitation, dynamic metrics (such as Computational Accuracy and Pass@$k$, used in 17 and 5 studies, respectively) have been incorporated to assess runtime behavior and execution correctness.
Beyond these traditional metrics, recent research has increasingly emphasized the importance of evaluating non-functional aspects of translated code, particularly in interactive development scenarios. These non-functional properties primarily include readability, extensibility, maintainability, and the cost of repair.

\begin{tcolorbox}[width=1.0\linewidth, title={}]
\textbf{RQ7: What evaluation metrics are used for neural code translation tasks?}
\begin{itemize}
    \item Neural code translation is evaluated using static and dynamic metrics. Among static metrics, BLEU is the most widely used, appearing in 42.1\% of studies, while among dynamic metrics, CA is the most prevalent, used in 29.8\% of studies.
    \item Beyond traditional metrics, recent research has placed greater emphasis on evaluating non-functional aspects of translated code, particularly in interactive development scenarios, focusing on readability, extensibility, and maintainability.
\end{itemize}
\end{tcolorbox}

%% file: sections/5.open.tex
\section{Open Issues and Opportunities}
\label{sec:open}

Despite significant progress in neural code translation research, several open issues remain, highlighting future opportunities and research directions. Addressing these challenges can lead to more robust, efficient, and secure neural code translation.

\textbf{Utilization of recent LLMs' techniques.}
In recent years, significant progress has been made in large language model technologies, particularly in enhancing model capabilities, interactivity, and adaptability to diverse application scenarios. Among the most active and promising directions are Retrieval Augmented Generation and LLM-based agents. However, the application of these technologies in neural code translation is still in its early stages and warrants further attention in future research.
Taking LLM-based agents~\cite{jin2024llms,he2024llm,liu2024large} as an example, in addition to designing agents such as the Initial Code Translator, Syntax Error Fixer, and Semantic Error Fixer~\cite{yuan2024transagent}, it is essential to develop additional agents like the Test Designer \& Executor, Code Reviewer \& Optimizer, and Feedback \& Iteration Agent. These agents automatically generate and execute test cases for the translated code to ensure its correctness. They also conduct code inspections to provide suggestions for improving readability, maintainability, and performance. Finally, based on the feedback, they guide other agents to make targeted corrections and optimizations.

\textbf{Code translation for low-resource programming languages.}
Low-resource programming languages refer to those with limited publicly available resources, such as large-scale code corpora, comprehensive documentation, and community or tool support~\cite{joel2024survey}. These languages, which include OCaml, Julia, and R, pose unique challenges in code translation tasks. 
Several promising research directions warrant further exploration for this type of code translation task. One direction involves training models on popular programming languages (such as Java or Python) and applying transfer learning techniques to adapt them to low-resource languages. A key challenge is to develop effective cross-lingual alignment mechanisms to bridge the semantic and syntactic differences between languages.
Another direction focuses on constructing synthetic parallel corpora through methods such as template-based translation, program transformation, and semantics-preserving code rewriting. These techniques can help overcome the scarcity of high-quality parallel data in low-resource settings.

\textbf{Security of code translation systems.}
The security of code language models is currently a popular research topic~\cite{chen2024security}. Neural code translation also faces security concerns. Here, we take backdoor attacks as an example for analysis.
Backdoor attacks on large language models refer to the intentional insertion of hidden malicious behaviors during the training or fine-tuning process, such that the model behaves normally under typical inputs but produces attacker-controlled outputs when triggered by specific inputs. These attacks can undermine the reliability and security of LLM-based systems. In the context of code translation, backdoor vulnerabilities pose serious challenges, such as silently injecting insecure code patterns, logic errors, or incorrect translations under trigger conditions, which can compromise software security and functionality. 
Research on backdoor attacks in neural code translation can be conducted from three perspectives: attack design, defense mechanisms, and evaluation frameworks.
From the attack perspective, different types of triggers can be designed based on syntax-level, semantic-level, and context-level to activate hidden malicious behaviors.
From the defense perspective, researchers can mitigate backdoor effects by fine-tuning models with clean data or by applying backdoor removal techniques such as fine-pruning or activation clustering to eliminate hidden behaviors.
From the evaluation framework perspective, it is valuable to construct standardized code translation datasets containing backdoor attack samples, which can be used to assess and improve the effectiveness of the aforementioned attack and defense techniques.

\textbf{Data leakage in neural code translation evaluation.}
In the context of large language models, data leakage refers to scenarios where test data has already been seen during pre-training. This leads the model to rely on memorization rather than demonstrating true generalization capabilities.
This issue is widespread in software engineering benchmarks~\cite{zhou2025lessleak}.
In neural code translation, data leakage also undermines the credibility of evaluation results by artificially inflating performance metrics and masking the model's true translation capabilities. This is especially critical in real-world scenarios, where a model's ability to generalize to unseen code is crucial.
To address the issue, one approach is to collect and use new data generated after pretraining large language models, which can effectively avoid the risk of data leakage. Additionally, the updated dataset should include a wider variety of code samples to provide a more reasonable evaluation of the model's actual performance.
On the other hand, data augmentation can also be used to alleviate this issue via semantic equivalence transformations or obfuscation techniques. Semantic equivalence transformation involves generating new code samples by applying semantic equivalence transformations to existing code. In contrast to semantic equivalence transformations, the goal of code obfuscation is not only to enhance the diversity of the dataset but primarily to increase the complexity of the code, especially by modifying the structure, names, and variables of the code, making it harder to understand or analyze.

\textbf{Repository-level neural code translation.}
Neural code translation can benefit significantly from incorporating repository-level information~\cite{wang2024reposvul}, as such context captures consistent coding styles, API usage patterns, naming conventions, and domain-specific logic that are often shared across files within the same repository. Leveraging this information can improve translation consistency, correctness, and readability. 
To effectively leverage repository-level information, several promising research directions can be explored.
First, it is important to develop repository-level graph representations that model the entire repository and its inter-file dependencies as structured graphs. Graph Neural Networks can then be employed to capture and propagate contextual information across files. Improving the efficiency of information propagation is critical for mitigating the impact of long-range dependencies and ensuring that the model can effectively utilize relationships spanning multiple files.
In addition, multimodal learning presents a valuable opportunity to enhance the model’s understanding of code by integrating various sources of information available in repositories, such as code, documentation, and comments.
Second, as code repositories grow in size, traditional neural models face increasing challenges related to computational complexity and memory consumption, especially when handling large-scale, multi-file projects. To address these issues, designing distributed training architectures is a promising direction. These architectures should support efficient parallel training on large repositories while preserving sensitivity to global repository context.
Moreover, incremental learning techniques can be explored to allow models to continuously learn from newly added code without the need for full retraining, thus improving scalability and adaptability in dynamic software development environments.

\textbf{Code repair and refinement for translated codes.}
In neural code translation, the translated code may contain errors or exhibit poor readability due to challenges in handling language differences, syntactic inconsistencies, or incomplete semantic understanding. As a result, code repair and code refinement become feasible methods to improve translation quality by fixing errors and enhancing the readability, maintainability, and performance of the output code. 
For translated code, future research can focus on improving quality from the perspectives of repair and refinement. 
In terms of repair, one promising direction is to develop intelligent strategies for locating and fixing errors by leveraging information from compiler diagnostics, runtime feedback, or static analysis. This can help automatically resolve common problems such as syntax errors, type mismatches, or incorrect API usage. 
In terms of refinement, the goal is to enhance code structure, naming, and control flow to improve readability and maintainability, all while preserving the original logic. Techniques such as program dependency analysis, data flow tracking, or support from external knowledge sources can be employed to guide this process. 
Another area of interest is to design unified models that combine code translation, code repair, and code refinement into a single workflow, where the model can generate code with built-in error correction and quality improvements. 
Finally, LLM–based code reviewer agents can be introduced to automatically inspect translated code, provide suggestions, and interact with developers to iteratively enhance code quality in practical development scenarios.

%% file: sections/6.threat.tex
\section{Threats to Validity}
\label{sec:threat}

In this section, we analyze the potential threats to the validity of our SLR.

\textbf{Incomplete literature search.} 
The search strategy may have missed some studies related to neural code translation, potentially affecting the completeness of our systematic literature review.
To mitigate this threat, we selected widely recognized digital libraries such as ACM Digital Library, ScienceDirect, IEEE Xplore, and Google Scholar, which host a substantial number of publications relevant to our keywords. Additionally, we designed a comprehensive search strategy that incorporated a broad set of keywords and applied the snowballing technique to maximize the inclusion of all related studies.

\textbf{ArXiv paper consideration.} 
The reason for selecting papers from arXiv is that research on neural code translation has progressed rapidly in recent years, while journals and conferences typically require a lengthy peer-review process. However, due to the absence of formal peer review, some arXiv papers may suffer from low quality. 
To mitigate this threat, we apply clearly defined quality assessment criteria, which help ensure that only high-quality arXiv papers are included in our study.

\textbf{Data extraction errors.} 
Errors may occur during the extraction of relevant information (such as code translation methods, evaluation metrics, or results), which could lead to inaccurate conclusions for certain research questions in neural code translation.
To alleviate this threat, we followed a standard data extraction protocol and two authors to cross-check and validate the extracted information to ensure its consistency and accuracy.

\textbf{Empirical knowledge bias. }
This systematic literature review covers {\primarystudies} studies on neural code translation and addresses seven research questions, requiring manual analysis of each study. Subjective judgment and empirical knowledge may introduce biases.  
To mitigate this, we took the following steps. 
First, as the first comprehensive review in this field, our goal was to provide an overview of the current state and trends in neural code translation. We referred to prior surveys on code generation~\cite{jiang2024survey,yang2023deep} to define research questions, focusing on task characteristics, pre-processing methods, code modeling, model construction, post-processing, evaluation subjects, and metrics. 
Second, to ensure accurate understanding, we extensively reviewed related literature before analyzing each study, predefining categories for each research question. This process improved analysis consistency and reduced the impact of empirical knowledge bias.

%% file: sections/7.conclusion.tex
\section{Conclusion and Future Work}
\label{sec:conclusion}

In this SLR, we review the progress of research in neural code translation.
Based on the collected {\primarystudies} primary studies, we systematically analyze existing work from seven aspects: task characteristics, pre-processing methods, code modeling methods, model construction methods, post-processing methods, evaluation subjects, and evaluation metrics. The analysis reveals that:
(1) Neural code translation mainly focuses on code translation between statically-typed languages (e.g., Java $\leftrightarrow$ C++), with a primary emphasis on function-level granularity;
(2) Data pre-processing (such as cleaning, deduplication, and augmentation) significantly improves data quality and diversity, while reducing compilation errors;
(3) Common code modeling methods include treating code as plain text, graph-based modeling, and conversion to intermediate representations;
(4) Fine-tuning is the most widely adopted model construction method, followed by training from scratch and prompt engineering. Emerging techniques, such as multi-agent collaboration and retrieval-augmented generation, are also being increasingly applied to code translation;
(5) Post-processing techniques are vital in neural code translation, improving correctness, readability, and alignment with target language conventions, with bug fixing being the most common;
(6) Frequently used evaluation subjects include TransCoder and CodeXGLUE, which are typically constructed based on open-source repository platforms such as GitHub;
(7) Neural code translation is evaluated using static metrics like BLEU and dynamic metrics like CA.

Based on the findings of our SLR, we have shown potential directions for future work from the perspectives of datasets, methodologies, and security, with a particular emphasis on leveraging the latest advancements in LLMs to further enhance code translation performance, considering more repository-level information, and applying repairs and refinements to improve the quality of the translated code.